\documentclass[12pt]{article}
\usepackage{amssymb,amsmath,latexsym,graphicx, mathrsfs, lscape, amsthm, bm,  hyperref}

\hypersetup{
    colorlinks=true,
    linkcolor=blue,
    citecolor=red,      
    urlcolor=cyan,
}

\numberwithin{equation}{section}

\begin{document}

    \thispagestyle{empty}

    \vspace{82pt}
    \begin{center}
        { \Large{\bf Taub-NUT as Bertrand spacetime with magnetic fields}}

        \vspace{8pt}

        { Sumanto Chanda $^\clubsuit$, Partha Guha$^\clubsuit$ and \ Raju Roychowdhury$^{\spadesuit}$}

        \vspace{8pt}

        {$\clubsuit$ \it S.N. Bose National Centre for Basic Sciences\\
        JD Block, Sector III, Salt Lake, Kolkata 700098, India \\
        \texttt{sumanto12@boson.bose.res.in, partha@bose.res.in}}

        {$\spadesuit$ \it Instituto de Fisica, Universidade de Sao Paulo,\\
                C. Postal 66318, 05314-970 Sao Paulo, SP, Brazil\\
        \texttt{raju@.if.usp.br, raju.roychowdhury@gmail.com}}

       \end{center}

\abstract{Based on symmetries Taub-NUT shares with Bertrand spacetime, we cast it as the latter 
with magnetic fields. Its nature as a Bianchi-IX gravitational instanton and other related 
geometrical properties are reviewed. We provide an easy derivation and comparison between 
the spatial Killing-Yano tensors deduced from first-integrals and the corresponding hyperk\"ahler 
structures and finally verify the existence of a graded Lie-algebra structure via Schouten-Nijenhuis 
brackets.}

\vspace{-0.5cm}
\tableofcontents

\setcounter{page}{1}
\section{Introduction}

The Taub-NUT \cite{extn} is an exact solution of Einstein's equations, found by Abraham 
Huskel Taub (1951), and extended to a larger manifold by E. Newman, T. Unti and L. Tamburino (1963). It 
is a gravitational anti-instanton with corresponding $SU(2)$ gauge fields, frequently studied for 
its geodesics which approximately describe the motion of well seperated monopole-monopole 
interactions. As a dynamical system it exhibits spherically symmetry, with geodesics admitting Kepler-type symmetry, implying 
first-integrals such as the angular momentum and Runge-Lenz vectors respectively. Witten's 
prescription \cite{witten} realized Taub-NUT space as a hyper-Kahler quotient using T-duality. This 
construction has a natural interpretation in terms of D-branes \cite{DM}, serving as an important 
example in string theory. 

The Bertrand spacetime metric, formulated by V. Perlick \cite{bert} is also spherically symmetric 

\vspace{-0.12cm}
\begin{equation}
\label{bst} ds^2 = h(\rho)^2 d\rho^2 + \rho^2 \big( d \theta^2 + \sin^2 \theta \ d \phi^2 \big) - \frac{dt^2}{\Gamma(\rho)}
\end{equation}

derived from Bertrand's Theorem, describing stable and closed geodesics with periodic orbits. Upon comparison, Euclidean Bertrand spaces and Taub-NUT spaces, appear quite similar apart from magentic monopole and dipole interaction of the Taub-NUT. This implies dynamical similarities, manifested through similar first-integrals characterizing their motion. It also implies that Taub-NUT possibly exhibits Kepler-Hooke configuration duality. 

Consequently, we try to find first-integrals similar to those associated with central-force motion under potentials involved in Bertrand's Theorem: the angular momentum and Laplace-Runge-Lenz vector. Since we are interested in the dynamical aspects of Taub-NUT spaces, our attention is directed toward geodesics and Killing tensors. Naturally, we will be looking at Killing tensors affiliated with Runge-Lenz-like vector. They obey the equation:

\vspace{-0.12cm}
\begin{equation}
\nabla_{(a} K_{b_1) b_2 . . . b_n} = 0
\end{equation}

Such tensors are the Killing-St$\ddot{\text{a}}$ckel tensors which are symmetric under index permutation and the Killing-Yano tensor. The Killing-Yano tensors are antisymmetric under index permutation, and their square gives the St$\ddot{\text{a}}$ckel tensor, like the antisymmetric tensor whose square gives the Runge-Lenz-like quantity as we shall see. Such Killing tensors exhibit quaternionic algebra, implying a connection to Hyperk$\ddot{\text{a}}$hler structures associated with the metric. \\

We start in section 2, with preliminaries on mechanical systems with magnetic field interactions, then compute first-integrals similar to the angular momentum and the Laplace-Runge-Lenz vector, in forms specifically for the Taub-NUT. We deduce such first-integrals using equations of motion and analytically using a momentum polynomial expansion. 

In section 3, we compare Taub-NUT metric to Euclidean Bertrand spacetime with magnetic monopoles and dipoles. Demonstrating such a similarity allows the intensely studied Bertrand spacetimes to share many important properties, and conversely extend properties of the Taub-NUT to Bertrand spaces with magnetic fields. This helps us identify symmetries and conserved quantities of Taub NUT and employ its curvature properties for Bertrand spacetimes. The last subsection covers the conserved quantity called the Fradkin tensor under Bohlin-Arnold-Vassiliev transformation which are bound to have such Killing tensors embedded. 

In section 4 we derive the Taub-NUT from a special case of self-dual Bianchi-IX metric described by the classical Darboux-Halphen system. Then we geometrically analyze it, computing curvature and confirming its self-duality as a gravitational instanton.

\vspace{-0.12cm}
\begin{equation}
R_{\mu \nu \rho \sigma} = \pm \frac12 {\varepsilon_{\mu \nu}}^{\lambda \gamma} R_{\lambda \gamma \rho \sigma}
\end{equation}

This analysis helps us explore the metric as an integrable system. Finally, we will compute topological invariants shared with comparable Bertrand spacetimes with magnetic fields.

In section 5, after a short introduction to Killing St$\ddot{\text{a}}$ckel tensor and Yano tensors, we will focus on the latter. After a brief overview of their properties, we will attempt to find them embedded within conserved quantities. Then, we see if it exhibits a graded Lie-algebra structure that decides if higher order Killing-Yano tensors can be constructed from it. 

Finally, in section 6 we derive hyperk$\ddot{\text{a}}$hler structures of the Taub-NUT. Then we compare them to the Killing-Yano tensors to see if they also exhibit quaternionic algebra.

In the last section we conclude our works and discuss the possibilities of further research 
along the line of the present article. The appendix contains detailed computation regarding Killing tensors following Holten´s algorithm, a brief review of Bohlin's transformation of the Taub-NUT and a double derivative expansion of Killing Yano tensors.

\numberwithin{equation}{section}

\section{Conserved Quantities}

In classical mechanics it is important to identify constants of motion called conserved quantities or first-integrals of the system. In the theory of integrable systems, all first-integrals are in involution or commute with each other within the Poisson Brackets, with at least one integral definitely being available. Given a $n+1$-spacetime metric with $t$ as cyclic variable:

\begin{equation}
ds^2 = g_{ij} (\bm{x}) dx^i dx^j + g_{tt} (\bm{x}) dt^2 + 2 g_{it} (\bm{x}) dx^i dt
\end{equation}

parameterized as $t = \tau$, we will have the Lagrangian and a conserved quantity $q$:

\vspace{-0.5cm}
\begin{equation}
L_{\dot{t} = 1} = \frac12 \big( g_{ij} (\bm{x}) \dot{x}^i \dot{x}^j \big) + \frac12 g_{tt} (\bm{x}) + g_{it} (\bm{x}) \dot{x}^i \hspace{1.5cm} q = \bigg( \frac{\partial L}{\partial \dot{t}} \bigg)_{\dot{t} = 1} = g_{tt} + g_{it} \dot{x}^i
\end{equation}

The Hamiltonian is given by the Legendre transform $H = \sum_{k \neq t} \frac{\partial L}{\partial \dot{x}^k} \dot{x}^k - L$, so that:

\begin{equation}
H = \frac12 g^{ij} (\bm{x}) \big( p_i - g_{it} (\bm{x}) \big) \big( p_j - g_{jt} (\bm{x}) \big) - \frac12 g_{tt} (\bm{x}) \hspace{1.5cm} p_i = \frac{\partial L}{\partial \dot{x}^i}
\end{equation}

In Hamiltonian dynamics, a conserved quantity $Q$ commutes with the Hamiltonian $H$, a first integral resulting from timetranslation invariance, within the Poisson Brackets:

\begin{equation}
\label{cons} \big\{ Q, H \big\} = 0
\end{equation}

However, this prescription is not gauge covariant for systems with gauge interactions. To better understand why, consider the following metric with scalar potential $U (\bm{x})$:

\begin{equation}
ds^2 = \delta_{ij} dx^i dx^j - \frac{ 1 + 2 U (\bm{x}) }m dt^2
\end{equation}

where $t$ is cyclical. Under the parameterization $t = \tau$, the lagrangian, Hamiltonian and Hamilton's dynamical equations for particles in presence of scalar potentials is given by:

\vspace{-0.25cm}
\begin{align}
L_{\dot{t} = 1} &= \frac m2 \bm{\dot{x}}^2 - U (\bm{x}) \\ 
H = \frac1{2m} \bm{p}^2 + U (\bm{x}) &\hspace{0.5cm} \Rightarrow \hspace{0.5cm}
\begin{cases}
\dot{\bold{x}} = \dfrac{\partial H}{\partial \bold{p}} = \dfrac{\bold{p}}m \\
\dot{\bold{p}} = - \dfrac{\partial H}{\partial \bold{x}} = - \bold{\nabla} U (\bm{x})
\end{cases}
\end{align}

For this system without magnetic fields, the fundamental brackets are:

\begin{equation}
\big\{ x^i, p_j \big\} = \delta^{ij} \hspace{1cm} \big\{ x^i, x^j \big\} = \big\{ p_i, p_j \big\} = 0
\end{equation}

Now, for charged particles in U(1) gauge fields from magnetic dipoles alone, without scalar potential, the metric is:

\begin{equation}
ds^2 = \delta_{ij} dx^i dx^j - \frac1m \big( dt^2 - 2 A_k (\bm{x}) dx^k dt \big)
\end{equation}

so the corresponding Lagrangian and Hamiltonian for $\dot{t} = 1$ are given by:

\vspace{-0.25cm}
\begin{equation}
\begin{split}
L_{\dot{t} = 1} &= \frac12 \Big( m \bm{\dot{x}}^2 - \dot{t}^2 + 2 \bm{A} ( \bm{x}) . \bm{\dot{x}} \dot{t} \Big) \hspace{1cm} q = \bigg( \frac{\partial L}{\partial \dot{t}} \bigg)_{\dot{t} = 1} = 1 - \bm{A} ( \bm{x}) . \bm{\dot{x}} \\
& \hspace{1cm} \therefore \hspace{1cm} H \approx \frac1{2m} \big( \bm{p} - \bm{A}(\bm{x}) \big)^2
\end{split}
\end{equation}

For charged particles in the presence of magnetic monopole and dipole U(1) gauge fields 
without scalar potential, the metric is:

\begin{equation}
\label{monomet} ds^2 = \delta_{ij} dx^i dx^j - \frac1m \big( d t - A_k (\bm{x}) dx^k \big)^2
\end{equation}

so the corresponding Lagrangian and Hamiltonian for $\dot{t} = 1$ are given by:

\vspace{-0.25cm}
\begin{equation}
\begin{split}
L_{\dot{t} = 1} &= \frac12 \Big[ m \bm{\dot{x}}^2 - \big( 1 - \bm{A} ( \bm{x}) . \bm{\dot{x}} \big)^2 \Big] \hspace{1.5cm} q = \bigg( \frac{\partial L}{\partial \dot{t}} \bigg)_{\dot{t} = 1} = 1 - \bm{A} ( \bm{x}) . \bm{\dot{x}} \\
& \hspace{1cm} \therefore \hspace{1cm} H = \frac1{2m} \big( \bold{p} - q \bm{A}( \bm{x}) \big)^2
\end{split}
\end{equation}

Now let us consider a Kaluza-Klein modification of this spacetime, such that we include another cyclical co-ordinate $\psi$ that is periodic along with magnetic field components coupled with it. This would result in a $4+1$ spacetime from a $3+1$ one given by:

\begin{equation}
ds^2 = \delta_{ij} dx^i dx^j + \frac1m \big( d \psi + A_k (\bm{x}) dx^k \big)^2 - \big( 1 + 2 V(\bm{x}) \big) dt^2
\end{equation}

so the Lagrangian and Hamiltonian for $\dot{t} = 1$, ignoring constant additive terms are:

\vspace{-0.25cm}
\begin{equation} \label{kklag}
\begin{split}
L &= \frac12 \Big[ m \bm{\dot{x}}^2 + \big( \dot{\psi} + \bm{A} ( \bm{x}) . \bm{\dot{x}} \big)^2 \Big] - V(r) \hspace{1.5cm} q = \frac{\partial L}{\partial \dot{\psi}} = \dot{\psi} - \bm{A} ( \bm{x}) \\
& \hspace{1cm} \therefore \hspace{1cm} H = \frac1{2m} \big( \bold{p} - q \bm{A}( \bm{x}) \big)^2 + V(r)
\end{split}
\end{equation}

where $q$ is a conserved charge. The corresponding Hamilton's equations are:

\begin{equation}
\dot{\bm{x}} = \frac{\partial H}{\partial \bm{p}} = \frac{\bm{p} - q \bm{A}}m \hspace{1.5cm} \dot{\bm{p}} = - \frac{\partial H}{\partial \bm{x}} = \frac qm \big( \bm{\nabla A} \big) . \big( \bm{p} - q \bm{A} \big) - \bm{\nabla} V
\end{equation}

Since the potentials are gauge dependent ($\bm{A} \rightarrow \bm{A} + \bm{\nabla} \Lambda$), the momenta therefore must be 
so as well ($\bm{p} \rightarrow \bm{p} + q \bm{\nabla} \Lambda$). Then, we must write gauge invariant momenta and express the 
Hamiltonian in its gauge invariant form. 

\begin{equation}
H = \frac{\bm{\Pi}^2}2 + V(r) \hspace{2cm} \bm{\Pi} = \bm{p} - q \bm{A}
\end{equation}

Any function and partial derivative operators in gauge invariant forms can be written as:

\vspace{-0.5cm}
\begin{equation}
\begin{split}
f(\bm{x}, &\bm{p}) \longrightarrow f(\bm{x}, \bm{\Pi}) \\
\frac{\partial \ }{\partial x^i} \hspace{0.25cm} &\longrightarrow \hspace{0.25cm} \frac{\partial \Pi^j}{\partial x^i} \frac{\partial \ }{\partial \Pi^j} + \frac{\partial \ }{\partial x^i} = - q \partial_i A_j \frac{\partial \ }{\partial \Pi^j} + \frac{\partial \ }{\partial x^i} \\
\frac{\partial \ }{\partial p^i} \hspace{0.25cm} &\longrightarrow \hspace{0.25cm} \frac{\partial \Pi^j}{\partial p^i} \frac{\partial \ }{\partial \Pi^j} + \frac{\partial \ }{\partial p^i} = \frac{\partial \ }{\partial \Pi^i} \hspace{1cm} \text{( No explicit  dependence on $\bm{p}$ )}
\end{split}
\end{equation}

with which the fundamental brackets become:

\begin{equation}
\big\{ x^i, \Pi_j \big\} = \delta^{ij} \hspace{1cm} \big\{ x^i, x^j \big\} = 0, \hspace{1cm} \big\{ \Pi_i, \Pi_j \big\} = - q F_{ij}
\end{equation}

where it is interesting to note that the new Poisson Brackets between the gauge covariant momenta are non-zero, as opposed to the usual case. This is a classical analogue of Ricci-identity (in the absence of torsion). We can furthermore redefine the Poisson Brackets as:

\begin{equation}
\big\{ f, g \big\} = \frac{\partial f}{\partial \bm{x}} \cdot \frac{\partial g}{\partial \bm{\Pi}} - \frac{\partial f}{\partial \bm{\Pi}} \cdot \frac{\partial g}{\partial \bm{x}} - q F_{ij} \frac{\partial f}{\partial \bm{\Pi}} \cdot \frac{\partial g}{\partial \bm{\Pi}}
\end{equation}

Now that we have redefined the Poisson Brackets to make Hamiltonian dynamics manifestly gauge invariant in the modified bracket, we can proceed to analyze the conserved quantities in a general gauge invariant form. This is done by the Holten Algorithm as shown in \cite{ggp} and \cite{crlv} discussed later as we shall see.

\numberwithin{equation}{subsection}

\subsection{A dynamical-systems description of Taub-NUT}

The Euclidean Taub-NUT metric as shown in \cite{extn} is given by:

\vspace{-0.5cm}
\begin{equation}
\begin{split}
\label{tn} ds^2 = f(r) \big\{ d r^2 &+ r^2 \big( d \theta^2 + \sin^2 \theta \ d\phi^2 \big) \big\} + g(r) \big( d \psi + \cos \theta d \phi \big)^2\\ 
\text{where} \hspace{1cm} &f(r) = 1 + \frac{4M}{r} \hspace{1cm} g(r) = \frac{(4M)^2}{ 1 + \frac{4M}{r} }
\end{split}
\end{equation}

For later reference, taking $d \widetilde{s}^2 = \dfrac{ds^2}{4M}$ we shall re-write the above metric into this form :

\vspace{-0.5cm}
\begin{equation}
\begin{split}
\label{altn} d \tilde{s}^2 &= V(r) \ \delta_{ij} \ d x^i d x^j + V^{-1} (r) \big( d \psi + \bm{A} . d \bm{x} \big)^2 \\ 
& \text{where} \hspace{1cm} V(r) = \frac1{4M} + \frac1{r} \hspace{1cm} \bm{A} . d \bm{x} = \cos \theta d \phi
\end{split}
\end{equation}

We now consider the geodesic flows of the generalized Taub-NUT metric given by (\ref{tn}), 
for which we can compose the Lagrangian:

\begin{equation}
\label{lag} \mathcal{L} = \frac12 f(r) \big\{ \dot{r}^2 + r^2 \big( \dot{\theta}^2 + \sin^2 \theta \ \dot{\phi}^2 \big) \big\} + \frac12 g(r) \big( \dot{\psi} + \cos \theta \dot{\phi} \big)^2
\end{equation}

We can further re-write the Lagrangian (\ref{lag}) into 3-dimensional form with a potential, 
as in (\ref{kklag}), independent of the $\psi$ as:

\begin{equation}
\mathcal{L} = \frac12 f(r) | \dot{\bm{x}} |^2 + \frac12 g(r) \big( \dot{\psi} + \bm{A} . \dot{\bm{x}} \big)^2 - U(r)
\end{equation}

where the momentum can be written as:

\begin{equation}
\bm{p} = \frac {\partial \mathcal{L}}{\partial \dot{\bm{x}}} = f(r) \dot{\bm{x}} + q \bm{A} \hspace{2cm} \bm{\Pi} = f(r) \dot{\bm{x}} = \bm{p} - q \bm{A}
\end{equation}

Spaces with the metric (\ref{tn}) exhibit $SU(2) \times U(1)$ isometry group. Given that we have at least 2 cyclical variables $\psi$ and $\phi$, we will have the following 4 Killing vectors given by:

\vspace{-0.5cm}
\begin{equation}
\begin{split}
D_0 &= \ \partial_\psi\\
D_1 &= - \sin \phi \ \partial_\theta - \cos \phi \cot \theta \ \partial_\phi + \frac{\cos \phi}{\sin \theta} \ \partial_\psi\\
D_2 &= \hspace{0.4cm} \cos \phi \ \partial_\theta - \sin \phi \cot \theta \ \partial_\phi + \frac{\sin \phi}{\sin \theta} \ \partial_\psi\\
D_3 &= \ \partial_\phi
\end{split}
\end{equation}

where $D_0$ commutes with all other killing vectors, while $D_1,D_2,D_3$ exhibit the $SU(2)$ Lie algebra given by $\big[ D_i, D_j \big] = - {\varepsilon_{ij}}^k D_k$. Since $\psi$ is cyclic, we have a conserved quantity:

\begin{equation}
q = \frac{\partial \mathcal{L}}{\partial \dot{\psi}} = g(r) \big( \dot{\psi} + \cos \theta \ \dot{\phi} \big) = g(r) \big( \dot{\psi} + \bm{A} . \dot{\bm{x}} \big) = const
\end{equation}

known as the relative electric charge. The symplectic 2-form $\omega$ and energy $\mathcal{E}$ for the Taub-
NUT system in Cartesian co-ordinates are:

\vspace{-0.5cm}
\begin{align}
\label{symp} \omega &= \frac12 \big( \omega_0 + q F (\bm{x}) \big)_{jk} dx^j \wedge dx^k = \sum_{i = 1}^3 d \big( p_i - q A_i (\bm{x}) \big) \wedge dx^i = \sum_{i = 1}^3 d \Pi_i \wedge dx^i \nonumber \\
& = \sum_{i = 1}^3 dp_i \wedge dx^i - \frac{q}{2 r^3} \sum_{i, j, k} \varepsilon_{ijk} x^i \ dx^j \wedge dx^k  \\
& \hspace{2cm} \mathcal{H} = \frac{| \bm{\Pi} |^2}{2 f(r)} + \frac{q^2}{2 g(r)} \hspace{1.5cm} F_{ij} (\bm{x}) = - \sum_k \varepsilon_{ijk} \frac {x^k}{r^3} 
\end{align}

Consequently, the Hamilton's equations are given by:

\vspace{-0.5cm}
\begin{equation}
\begin{split}
\dot{\bm x} = \big\{ \bm{x}, \mathcal{H} \big\}_\theta &= \frac{\bm \Pi}{f(r)} \hspace{1.5cm} \dot{\bm \Pi} = \big\{ \bm{\Pi}, \mathcal{H} \big\}_\theta = \alpha (r) \frac{\bm x}r + \frac q{r^3 f (r)} \bm{x} \times \bm{\Pi} - \bm{\nabla} U (r) \\
\text{where} &\hspace{1cm} \alpha (r) = \frac{f'(r)}{2 \big( f(r) \big)^2} \big| \bm{\Pi} \big|^2 + \frac{g'(r)}{2 \big( g(r) \big)^2}
\end{split}
\end{equation}

Using these equations, we find angular momentum in presence of magnetic fields to be:

\vspace{-0.35cm}
\[ \bigg( \frac{d \bm{x} }{d t} \times \bm{\Pi} + \bm{x} \times \frac{d \bm{\Pi}}{d t} \bigg) = \frac q{r^3 f(r)} \big[ \bm{x} \times \big( \bm{x} \times \bm{\Pi} \big) \big] = q \bigg[ \frac{ \big( \bm{x} . \bm{\dot{x}} \big) \bm{x}}{r^3} - \frac{\bm{\dot{x}}}r \bigg] = - q \frac{d \ }{d t} \bigg( \frac{\bm{x}}r \bigg) \]

\begin{equation}
\label{angmom} \therefore \hspace{1cm} \frac{d \ }{d t} \bigg( \bm{x} \times \bm{\Pi} + q \frac{\bm{x}}r \bigg) = 0 \hspace{1cm} \Rightarrow \hspace{1cm} \bm{J} = \bm{x} \times \bm{\Pi} + q \frac{\bm{x}}r
\end{equation}

The cyclic variable allows reduction of the geodesic flow on $T(\mathbb{R}^4 - \{ 0 \})$ to a system on $T(\mathbb{R}^3 - \{ 0 \})$. The reduced system's rotational invariance implies it must have a conserved energy, angular momentum and vector $\bm{K}$ analogous to the Laplace-Runge-Lenz vector:

\vspace{-0.35cm}
\begin{align}
\label{tham} \mathcal{H} &= \frac12 \frac{\bm{\Pi}^2}{f(r)} + \bigg( \frac12 \frac{q^2}{g(r)} + U(r) \bigg) = \frac12 \frac{\bm{\Pi}^2}{f(r)} + W(r) \\
\bm{J} &= \bm{x} \times \bm{\Pi} + q \frac{\bm{x}}{r} \\
\bm{K} &= \frac12 K_{\mu \nu} \dot{x}^\mu \dot{x}^\nu = \bm{\Pi} \times \bm{J} + \bigg( \frac{q^2}{4m} - 4mE \bigg) \frac{\bm{x}}r
\end{align}

This concludes the detailing of conserved quantities of the Taub-NUT from a dynamical systems perspective. Now we shall proceed to consider a systematic analytic process that describes conserved quantities as power series expansions of momenta.

\subsection{Holten Algorithm description}

One way of analytically obtaining conserved quantities that are polynomials in momenta is by writing them in a power series expansion involving the gauge invariant momenta:

\vspace{-0.125cm}
\begin{equation}
Q = C^{(0)} (\bold{r}) + C_i^{(1)} (\bold{r}) \Pi^i + \frac1{2!} C_{ij}^{(2)} (\bold{r}) \Pi^i \Pi^j + \frac1{3!} C_{ijk}^{(3)} (\bold{r}) \Pi^i \Pi^j \Pi^k + . . . . 
\end{equation}

where all the coefficients of momenta power series are symmetric under index permutation. Applying this to eq (\ref{cons}), we can obtain the relations for each coefficient by matching the appropriate product series of momenta for both the terms. 

\vspace{-0.125cm}
\[ \big\{ Q, H \big\} = \sum_n \Big[ \big\{ C^{(n)}_{ \{ i \} } \prod_{ \{i\} } \Pi^k , \Pi^j \big\} \Pi_j + \big\{ C^{(n)}_{ \{ i \} } \prod_{ \{i\} } \Pi^k, V(r) \big\} \Big]= 0 \]

\begin{equation}
\therefore \hspace{1cm} \nabla_j C^{(n)}_{ \{ m \} } \prod_{\{ m \}} \Pi^k = q C^{(n+1)}_{ \{ m \} i } \Big( F^{ij} + \partial_j V(r) \Big) \prod_{ ( \{m\}, k \neq i) } \Pi^k
\end{equation}

The equations we will get up to the 3rd order setting $C^{(i)}_{ \{m\} } = 0 \hspace{0.25cm} \forall \hspace{0.25cm} i \geq 3$ are:

\vspace{-0.5cm}
\begin{equation}
\begin{split}
\label{odr2} \text{order 0:} & \hspace{1cm}0 = C^{(1)}_m \partial_m \big( V(r) \big) \\
\text{order 1:} &\hspace{1cm} \nabla_i C^{(0)} = q F_{ij} C^{(1)}_j + C^{(2)}_{ij} \partial_j \big( V(r) \big) \\
\text{order 2:} &\hspace{1cm} \nabla_i C^{(1)}_j + \nabla_j C^{(1)}_i = q \big( F_{im} {C^{(2)m}}_j + F_{jm} {C^{(2)m}}_i \big) \\
\text{order 3:} &\hspace{1cm} \nabla_i C^{(2)}_{jk} + \nabla_k C^{(2)}_{ij} + \nabla_j C^{(2)}_{ki} = 0
\end{split}
\end{equation}

Now we will turn our attention to some familiar conserved quantities.

\subsubsection{Some basic Killing Tensors}

Using the above relations for the various terms, we will now look at some familiar ones that we have already studied in classical mechanics. 

\subsubsection*{Angular Momentum}

The conserved quantity that results from the 1st order term of the Holten series alone is:

\vspace{-0.5cm}
\[ \begin{split}
Q^{(1)} &= C_i^{(1)} \Pi^i = - g_{im} (\vec{x}) {\varepsilon^m}_{jk} \theta^k x^j \Pi^i \\ 
\Rightarrow \hspace{1cm} & \ \bm{L} . \bm{\theta} = - \big( \varepsilon_{ijk} \Pi^i x^j \big) \theta^k = \big( \bm{x} \times \bm{\Pi} \big) . \bm{\theta}
\end{split} \]

\begin{equation}
\therefore \hspace{1cm} \bm{L} = \bm{x} \times \bm{\Pi}
\end{equation}

This eventually becomes the conserved quantity known as the angular momentum.

\subsubsection*{Laplace-Runge-Lenz vector}

On the other hand, the conserved quantity from the 2nd order term of the series alone is:

\vspace{-0.5cm}
\[ \begin{split}
Q^{(2)} &= \frac12 C^{(2)}_{ij} \Pi^i \Pi^j = \big\{ \big| \bm{\Pi} \big|^2 \big( \bm{n} . \bm{x} \big) - \big( \bm{\Pi} . \bm{x} \big) \big( \bm{\Pi} . \bm{n} \big) \big\} \\
\Rightarrow \hspace{1cm} & \ \bm{N} . \bm{n} = \big\{ \big| \bm{\Pi} \big|^2 \bm{x} - \big( \bm{\Pi} . \bm{x} \big) \bm{\Pi} \big\} . \bm{n} = \big\{ \bm{\Pi} \times \big( \bm{x} \times \bm{\Pi} \big) \big\} . \bm{n}
\end{split} \]

\begin{equation}
\therefore \hspace{1cm} \bm{N} = \bm{\Pi} \times \big( \bm{x} \times \bm{\Pi} \big)
\end{equation}

This quantity is a term contained in another conserved quantity known as the Laplace-Runge-Lenz vector. Having found the two familiar types of conserved quantities, we can now proceed to see what it looks like for the Taub-NUT metric.

\subsubsection{Holten algorithm for Taub-NUT}

Now, for the  Taub-NUT metric, we have (\ref{tham}) giving the Hamiltonian. This can be written 
in dimensionally reduced form as:

\vspace{-0.25cm}
\begin{equation}
\mathcal{H} = \frac12 | \bm{\Pi} |^2 + f(r) W(r) \hspace{2cm} W(r) = U(r) + \frac{q^2}{2 h(r)} + \frac{\mathcal{E}}{f(r)} - \mathcal{E}
\end{equation}

From this Hamiltonian, after setting all higher orders $C^{(2)}_{ij} = C^{(3)}_{ijk} = 0$, we get the modified 
1st and 2nd order equations to be the following:

\vspace{-0.5cm}
\begin{equation}
\begin{split}
\label{odr1} \text{order 1:} &\hspace{1cm} \partial_i C^{(0)} = q F_{ij} C^{(1)j} \\
\text{order 2:} &\hspace{1cm} \nabla_i C^{(1)}_j + \nabla_j C^{(1)}_i = 0
\end{split}
\end{equation}

The constraint equation of the 2nd order of (\ref{odr1}) gives us:

\vspace{-0.25cm}
\begin{equation}
C^{(1)}_i = g_{im} (\bm{x}) {\varepsilon^m}_{jk} \theta^j x^k
\end{equation}

\vspace{-0.25cm}
\[ \partial_i C^{(0)} = \frac{q}{r^3} \varepsilon_{ijk} \ {\varepsilon^j}_{nm} x^k \theta^m x^n \hspace{0.25cm} \equiv \hspace{0.25cm} \frac{q}{r^3} \big[ \bm{x} \times \big( \bm{\theta} \times \bm{x} \big) \big]_i = \frac{q}{r^3} \big[ r^2 \bm{\theta} - \big( \bm{x} . \bm{\theta} \big) \bm{x} \big]_i \]

\vspace{-0.25cm}
\begin{equation}
\therefore \hspace{1cm} \nabla_i C^{(0)} = q \bigg( \frac{\theta_i}r - \frac{\big( \bm{x} . \bm{\theta} \big) x_i}{r^3} \bigg) \hspace{1cm} \Rightarrow \hspace{1cm} C^{(0)} = q \theta_i \frac{x^i}r
\end{equation}

Thus, we have the overall solution, and the corresponding conserved quantity:

\vspace{-0.5cm}
\begin{align}
\label{amom} Q \equiv J_k \theta^k = C^{(0)} + C^{(1)}_i \Pi^i &= \bigg( - g_{im} (\vec{x}) {\varepsilon^m}_{jk} x^j \Pi^i + q \frac{x_k}r \bigg) \theta^k \\ 
\therefore \hspace{1cm} \bm{J} . \bm{\theta} = \bigg( \bm{x} \times \bm{\Pi} + q \frac{\bm{x}}r \bigg) . \bm{\theta} &\hspace{1cm} \Rightarrow \hspace{1cm}  \bm{J} = \bm{x} \times \bm{\Pi} + q \frac{\bm{x}}r 
\end{align}

However, if we explore upto the 2nd order, setting $C^{(2)}_{ij} \neq 0$, we will return to the equations 
in (\ref{odr2}). For the 3rd order equation, the solution for $C^{(2)}_{ij}$ is given by (\ref{con2}), so that:

\begin{equation}
C^{(2)}_{ij} = \big( 2 g_{ij} (\bm{x}) n_k - g_{ik} (\bm{x}) n_j - g_{kj} (\bm{x}) n_i \big) x^k
\end{equation}

Eventually the other co-efficients are given by:

\vspace{-0.25cm}
\begin{equation}
\nabla_{(i} C^{(1)}_{j)} = q \big( F_{ik} C^{(2)}_{kj} + F_{jk} C^{(2)}_{ki} \big)
\end{equation}

\vspace{-0.5cm}
\[ \begin{split}
F_{ik} C^{(2)}_{kj} &= - 2 \varepsilon_{ijn} \frac {x^n}{r^3} \underbrace{ \big( n_m x^m \big)}_{\bm{n} . \bm{x}} + \underbrace{\varepsilon_{ikn} \frac {x^k x^n}{r^3} n_j}_0 + \ \underbrace{ \varepsilon_{ikn} n^k x^n}_{( \bm{n} \times \bm{x} )_i} \ \frac{x_j}{r^3} \\ 
\therefore \hspace{1cm} \nabla_i &C^{(1)}_j + \nabla_j C^{(1)}_i = q \bigg\{ \frac {x_j}{r^3} ( \bm{n} \times \bm{x} )_i + \frac {x_i}{r^3} ( \bm{n} \times \bm{x} )_j \bigg\}
\end{split} \]

Here, one can choose to insert extra terms:

\vspace{-0.125cm}
\[ \nabla_i C^{(1)}_j + \nabla_j C^{(1)}_i = - q \bigg\{ \nabla_j \bigg( \frac {\varepsilon_{ikm} n^k x^m}r \bigg) + \nabla_i \bigg( \frac {\varepsilon_{jkm} n^k x^m}r \bigg) \bigg\} \]

Thus, we can easily see which term on the RHS corresponds to what on the LHS, allowing 
us to solve for the 1st order and zeroth order coefficients  from (\ref{odr2}) :

\vspace{-0.125cm}
\begin{equation}
C^{(1)}_i = - \frac{q}r g_{im} (\bm{x}) {\varepsilon^m}_{jk} n^k x^j
\end{equation}

\vspace{-0.25cm}
\[ \nabla_i C^{(0)} = - q^2 \bigg( \frac{n_i}{r^2} - \frac{\big( \bm{x} . \bm{n} \big) x_i}{r^4} \bigg) + \big\{ 2 \big( \bm{n} . \bm{x} \big) \delta_{ij} - n_i x_j - x_i n_j \big\} \partial_j\bigg( f(r)U(r) + q^2 \frac{f(r)}{2 g(r)} + \mathcal{E} - \mathcal{E} f(r) \bigg) \]

In the case of the generalised Taub-NUT metric, the most general potentials admitting a 
Runge-Lenz vector are of the form:

\vspace{-0.25cm}
\begin{equation}
U(r) = \frac1{f(r)} \bigg( \frac{q^2}{2 r^2} + \frac{\beta}r + \gamma \bigg) - \frac{q^2}{2 g(r)} + \mathcal{E}
\end{equation}

\begin{equation}
\nabla_i C^{(0)} = \beta \bigg( \frac{n_i}r - \frac{\big( \bm{n} . \bm{x} \big) x_i}{r^3} \bigg) \hspace{1.5cm} C^{(0)} = \beta n_i \frac{x^i}r
\end{equation}

For integrability, we require the commutation relation:

\vspace{-0.5cm}
\begin{equation}
\begin{split}
\big[ \partial_i, \partial_j \big] C^{(0)} = 0 \hspace{0.5cm} &\Rightarrow \hspace{0.5cm} \Delta \bigg( f(r) W(r) - \frac{q^2 g^2}{2 r^2} \bigg) = 0 \\
\Rightarrow \hspace{0.5cm} f(r) W(r) - \frac{q^2 g^2}{2 r^2} = \frac{\beta}r + \gamma \hspace{0.5cm} &\Rightarrow \hspace{0.5cm} f(r) W(r) = \frac{q^2 g^2}{2 r^2} + \frac{\beta}r + \gamma \hspace{1cm} \beta, \gamma \in \mathbb{R}
\end{split}
\end{equation}

Thus, this overall conserved quantity is given as:

\vspace{-0.5cm}
\begin{align}
Q \equiv R_k \theta^k = C^{(0)} + C^{(1)}_i \Pi^i &+ C^{(2)}_{ij} \Pi^i \Pi^j \nonumber \\ 
\therefore \hspace{1cm} \bm{R} . \bm{n} = \bigg( \bm{\Pi} \times \big( \bm{x} \times \bm{\Pi} \big) - \frac{q}r \bm{x} \times \bm{\Pi} + \beta \frac{\bm{x}}r \bigg) . \bm{n} \hspace{0.5cm} &\Rightarrow \hspace{0.5cm} \bm{R} = \bm{\Pi} \times \bm{J} + \beta \frac{\bm{x}}r
\end{align}

Now we will take a detour to look at some details regarding the Runge-Lenz vector. 

\section{Bertrand spacetime dualities}

In Newtonian mechanics, there are only two potentials allowing stable, closed and periodic 
orbits: Hooke's Oscillator ($V(r) = a r^2 + b$), and Kepler's Orbital Motion ($\Gamma(r) = \frac ar + b$) 
potentials. There is a relativistic analogue, given by the corresponding metrics in \cite{bert}, describing 
spherically symmetric and static spacetime, with bounded and periodic trajectories. The Taub-
NUT is one example of a spherically symmetric spacetime. Naturally, one would ask how it 
compares with the Euclidean Bertrand spacetime (BST) metric with magnetic fields. 

\subsection{Bertrand spacetimes with magnetic fields}

The Bertrand spacetime metric is given by (\ref{bst}). If we take the Euclidean version and include 
magnetic monopole and dipole interaction terms, then the metric becomes like (\ref{monomet}) as:

\begin{equation}
ds^2 = h(\rho)^2 d\rho^2 + \rho^2 \big( d \theta^2 + \sin^2 \theta \ d \phi^2 \big) + \frac1{\Gamma(\rho)} \big( d t + A_i d x^i \big)^2
\end{equation}

If we recall, the Taub-NUT metric was given by (\ref{tn}). To see how they are comparable, 
we shall attempt a co-ordinate map.

\vspace{-0.5cm}
\[ \begin{split}
f(r) \ d r^2 &= h(\rho)^2 d\rho^2 \hspace{1cm} f(r) r^2 = \rho^2 \hspace{1cm} g(r)  = \frac 1{\Gamma(\rho)} \hspace{1cm} t = \psi + k\\ 
\Rightarrow \hspace{1cm} \frac{d r}{r} &= \frac{h(\rho) \ d\rho}\rho \hspace{1cm} \Rightarrow \hspace{1cm} r = r_0 \ e^{\int d \rho \ \frac{h(\rho)}\rho}
\end{split} \]

Thus, we can suppose that Taub-NUT metric resembles Bertrand spacetime with magnetic 
fields. We can also proceed the other way around, starting with the generalized Taub-NUT 
metric and proceeding toward Bertrand spacetimes by applying appropriate potential power 
laws shown in \cite{dynsym}. Thus,  like the BSTs, there are two Taub-NUT configurations:

\begin{enumerate}

\item Hooke's Oscillator configuration: 

\vspace{-0.25cm}
\[ f_{\mathcal{O}}(r) = a r^2 + b \hspace{1cm} g_{\mathcal{O}}(r) = \dfrac{ r^2 \big( a r^2 + b \big)}{c r^4 + d r^2 + 1} \]

\item Kepler's orbital configuration: 

\vspace{-0.25cm}
\[ f_{\mathcal{K}}(r) = \dfrac{a + br}r \hspace{1cm} g_{\mathcal{K}}(r) = \dfrac{r \big( a + b r \big)}{c r^2 + d r + 1} \]

\end{enumerate}

Evidence for the duality between these two configurations of the metric can be clearly 
demonstrated. To study Taub-NUT space duality, we confine motion to a cone ($\theta = const$). 
This is permissible because of the conserved angular momentum (\ref{angmom}), for which \cite{extn, dynsym, js}

\vspace{-0.12cm}
\begin{equation}
\label{cone} \bm{J} . \bm{e^r} = \big| \bm{J} \big| \cos \theta = const \hspace{1cm} \Rightarrow \hspace{1cm} \theta = const
\end{equation}

This allows us to reduce the problem to 2-dimensions by rendering $\theta$ a constant co-ordinate, 
allowing us to write the metric as:

\vspace{-0.5cm}
\begin{equation}
ds^2 = f(r) \big( d r^2 + r^2 \alpha^2 \ d \phi^2 \big) + g(r) \big( d \psi + \beta \ d \phi \big)^2 \hspace{1cm} \alpha = \sin \theta, \beta = \cos \theta
\end{equation}

We shall represent the co-ordinates as $Z = x + i y, \ \xi = X + i Y$, where $|Z| = r \cos \frac{\theta}2$ and perform Bohlin's 
transformation \cite{Bohlin} of the Oscillator metric ($Z \rightarrow \xi = Z^2$). This requires the complex co-ordinates defined for self-dual Euclidean spaces \cite{eh}, where ($\theta = const$) (\ref{cone}):

\vspace{-0.5cm}
\begin{align}
Z = x &+ i y = |Z| \exp \bigg[ \frac i2 \big( \psi + \phi \big) \bigg] \hspace{1.5cm} \xi = X + i Y = | \xi | \exp \bigg[ \frac i2 \big( \chi + \Phi \big) \bigg] \\
Z \rightarrow \xi &= Z^2 = |Z|^2 \exp \big[ i \big( \psi + \phi \big) \big] \hspace{1cm} \Rightarrow \hspace{1cm} \phi \rightarrow \Phi = 2 \phi, \ \psi \rightarrow \chi = 2 \psi
\end{align}

\begin{equation}
\big( ds^2 \big)_\mathcal{O} = \big( a |Z|^2 + b \big) |d Z|^2 + \frac{|Z|^2 (a |Z|^2 + b)}{c |Z|^4 + d|Z|^2 + 1} \big( d \psi + \beta \ d \phi \big)^2
\end{equation}

\[ \begin{split}
\big( a |Z|^2 + b \big) |d Z|^2 &+ \frac{|Z|^2 (a |Z|^2 + b)}{c |Z|^4 + d|Z|^2 + 1} \big( d \psi + \beta \ d \phi \big)^2 \\
& \hspace{-1.5cm} Z \rightarrow \xi = Z^2 \bigg{\updownarrow} \phi \rightarrow \Phi = 2 \phi, \ \psi \rightarrow \chi = 2 \psi \\
 \frac14 \bigg\{ \frac{a |\xi| + b}{|\xi|} |d \xi|^2 &+ \frac{|\xi| \big( a |\xi| + b \big)}{c|\xi|^2 + d|\xi| + 1} \big( d \chi + \beta \ d \Phi \big)^2 \bigg\} 
\end{split} \]

Then we can compare with the Kepler system in presence of magnetic fields:

\begin{equation}
\big( ds^2 \big)_\mathcal{K} = \frac{b |Z| + a}{|Z|} |d Z|^2 + \frac{|Z| \big( b |Z| + a \big)}{c|Z|^2 + d|Z| + 1} \big( d \chi + \beta \ d \Phi \big)^2
\end{equation}

showing that aside from a factor of $\dfrac14$, a variable swap $a \leftrightarrow b$ completes the transformation, 
and thus, the two configurations of Taub-NUT are also related via Bohlin's transformation like 
Bertrand spacetime. For various settings of the constants, one can get different configurations 
of spacetime, as shall be described in the following table.

\begin{center}
\textbf{Systems for various settings}\\
\begin{tabular}{|c|c|c|c|c|c|c|c|}

\hline & & & & & & &\\

\textbf{Type} & $\bm{a}$ & $ \ \bm{b} \ $ & $ \ \bm{c} \ $ & $ \ \bm{d} \ $ & $\bm{f(r)}$ & $\bm{g(r)}$ & \textbf{System Name} \\ & & & & & & &\\

\hline & & & & & & &\\

$\mathcal{K}$
&
$0$
&
$1$
&
$1$
&
$- 2$
&
$1$
&
$\dfrac{r^2}{(1 - r)^2}$
&
MIC-Zwangier\\ & & & & & & &\\

\hline & & & & & & &\\

$\mathcal{K}$
&
$0$
&
$1$
&
$0$
&
$- \dfrac{2k}{q^2}$
&
$1$
&
$\dfrac{r^2}{1 - \frac{2k}{q^2} r}$
&
MIC-Kepler\\ & & & & & & &\\

\hline & & & & & & &\\

$\mathcal{O}$
&
$0$
&
$1$
&
$\dfrac{k}{q^2}$
&
$0$
&
$1$
&
$\dfrac{r^2}{1 + \frac{k}{q^2} r^4}$
&
MIC-Oscillator\\ & & & & & & &\\

\hline & & & & & & &\\

$\mathcal{K}$
&
$4m$
&
$1$
&
$0$
&
$\dfrac1{4m}$
&
$\dfrac{4m + r}r$
&
$\dfrac{(4m)^2 r}{4m + r}$
&
Euclidean Taub-NUT\\

& & & & & & &\\

\hline
\end{tabular}\\ \vspace{0.5cm}
$\mathcal{K}$ - Kepler, \hspace{1cm} $\mathcal{O}$ - Oscillator
\end{center}

\subsection{Kepler-Oscillator duality}

In the study of central force problem, we learn that the Kepler and Oscillator systems are dual 
to each other according a duality map demonstrated in \cite{bohlin} and \cite{dual}. This is summed up in 
Bertrand's theorem which describes them as the only systems with stable, closed and periodic 
orbits. Thus, curved Bertrand space-times are classified as Type I and Type II, representing 
Kepler and Oscillator systems respectively. 

If we start with the 2-dimensional simple harmonic oscillator described by $\ddot{x}^i = - \omega^2 x^i$, we 
are reminded of a conserved tensorial quantity, known as the Fradkin tensor:

\begin{equation}
T^{ij} = p^i p^j + \kappa x^i x^j \hspace{2cm} i, j = 1, 2
\end{equation}

Any conserved quantity can be obtained by contracting the Fradkin tensor over its two 
indices by any chosen structure. ie.

\vspace{-0.25cm}
\begin{equation}
Q = M_{ij} T^{ij}
\end{equation}

This quantity is symmetric under index permutation. Its complex counterpart is given by:

\vspace{-0.5cm}
\begin{align}
T_{z^a z^b} &= G^{ij}_{z^a z^b} T_{ij} \hspace{2cm} z^a = \big\{ z, \bar{z} \big\}\\ 
G_{z z} = \left( {\begin{array}{cc} 1 & i \\ i & -1 \end{array}} \right) \hspace{1cm} 
G_{\bar{z} \bar{z}} &= \left( {\begin{array}{cc} 1 & -i \\ -i & -1 \end{array}} \right) \hspace{1cm} 
G_{z \bar{z}} = G_{\bar{z} z} = \left( {\begin{array}{cc} 1 & 0 \\ 0 & 1 \end{array}} \right)
\end{align}

According to the Arnold-Vasiliev duality \cite{AV}, a co-ordinate transformation and re-parametrization 
of the first two complex Fradkin tensors will give us the Laplace-Runge-Lenz vector. 

\begin{equation}
\bm{A} = \bm{p} \times \bm{L} + \beta \frac{\bm{x}}r
\end{equation}

In tensorial form, this is written as follows:

\begin{equation}
A_i = \varepsilon_{ikl} {\varepsilon^l}_{jm} p^k x^j p^m + \frac{\beta}r \delta_{ij} x^j = x^j \Big\{ \big( \delta_{ij} \delta_{km} - \delta_{ik} \delta_{jm} \big) p^k p^m + \frac{\beta}r \delta_{ij} \Big\}
\end{equation}

showing that the 1st term can be expressed in a form quadratic in mommenta. Since it 
is essentially a linear combination of Fradkin tensor components, we would prefer it to be 
symmetric in the momentum indices like its oscillator counterpart. Thus, we can write

\begin{equation}
A_i = x^j \bigg\{ \frac12 \big( 2 \delta_{ij} \delta_{km} - \delta_{ik} \delta_{jm} - \delta_{im} \delta_{jk} \big) p^k p^m + \frac{\beta}r \delta_{ij} \bigg\}
\end{equation}

Hence, to describe this conserved quantity of the Kepler system, we need tensors that are:

\begin{enumerate}

\item quadratic in momenta

\item symmetric under index permutation

\item conserved along geodesics

\end{enumerate}

Our next step will be to explore such tensors in the next section.

\numberwithin{equation}{section}

\section{A review of geometric properties}

An instanton or pseudo-particle is a concept in mathematical physics that describes solutions 
to equations of motion of classical field theory on a Euclidean spacetime. The first such 
solutions discovered were found to be localized in spacetime, hence, the name instanton or 
pseudoparticle. Instantons are important  in quantum field theory because:

\begin{enumerate}

\item They are leading quantum corrections to classical motion equations in the path integral

\item They are useful for studying tunneling behaviour in systems like the Yang-Mills theory

\end{enumerate}

 Since we are considering the Taub-NUT metric defined on a 4-dimensional Euclidean plane, 
it is worthwhile to verify if it is an instanton as well. In this section, we will proceed to analyze 
its geometrical properties exhaustively, verify if the Taub-NUT is an instanton from the curvature 
components computed from the metric, and also take a look at its topological properties.\\

A variable transformation $m = 2 M$ and $\rho =  r - 2 M$ of (\ref{tn}) gives the Taub-NUT as:

\vspace{-0.5cm}
\begin{equation}
ds^2 = \frac{r + m}{r - m} dr^2 + 4m^2 \frac{r - m}{r + m} \big( d \psi + \cos \theta \ d \phi \big)^2 + \big( r^2 - m^2 \big) \big( d \theta^2 + \sin^2 \theta \ d\phi^2 \big)
\end{equation}

This can be further recast into the form:

\begin{equation}
\label{tnuteq} ds^2 = \frac{r + m}{r - m} dr^2 + 4m^2 \frac{r - m}{r + m} \sigma_1^2 + \big( r^2 - m^2 \big) \big( \sigma_2^2 + \sigma_3^2 \big) 
\end{equation}

where the variables $\sigma_i$ are essentially solid angle elements in 4-dimensional Euclidean space 
obeying the following structure equation:

\vspace{-0.25cm}
\begin{equation}
\label{struc} d \sigma^i = - {\varepsilon^i}_{jk} \ \sigma^j \wedge \sigma^k \hspace{2cm} \sigma^i = - \frac1{r^2} \eta^i_{\mu \nu} x^\mu dx^\nu
\end{equation}

Now that we have identified the vierbeins, we will proceed to implement Cartan's  method 
of computing spin connections and the Riemann curvature components. Embedded within 
them are the $SU(2)$ gauge fields and their corresponding field strengths as we shall see.

\numberwithin{equation}{subsection}

\subsection{Taub-NUT as a Darboux-Halphen system}

The Taub-NUT is a special case of self-dual Bianchi-IX metrics \cite{cgr1}, which are characterized 
by the classical Darboux-Halphen system. The self-dual metric and its characteristic system 
of equations are given by:

\vspace{-0.5cm}
\begin{align}
\label{bianchi} d \widetilde{s}^2 = \big( \Omega_1 \Omega_2 \Omega_3 \big) & d \widetilde{r}^2 + \frac{\Omega_2 \Omega_3}{\Omega_1} \big( \sigma_1 \big)^2 + \frac{\Omega_3 \Omega_1}{\Omega_2} \big( \sigma_2 \big)^2 + \frac{\Omega_1 \Omega_2}{\Omega_3} \big( \sigma_3 \big)^2 \\ \nonumber \\
\begin{split}
\Omega'_1 &= \Omega_2 \Omega_3 - \Omega_1 \big( \Omega_2 + \Omega_3 \big) \\
\Omega'_2 &= \Omega_3 \Omega_1 - \Omega_2 \big( \Omega_3 + \Omega_1 \big) \\
\Omega'_3 &= \Omega_1 \Omega_2 - \Omega_3 \big( \Omega_1 + \Omega_2 \big)
\end{split} \hspace{1.5cm} ( \ )' = \frac{d \ }{d \widetilde{r}} ( \ )
\end{align}

where $\Omega_i$ are parameters defined to re-write the Bianchi-IX metric into the above form 
(\ref{bianchi}) to write self-dual equations. One particular first integral of the this system \cite{cgr2} is:

\vspace{-0.25cm}
\begin{equation}
Q = \frac{(\Omega_1)^2}{(\Omega_3 - \Omega_1) (\Omega_1 - \Omega_2)} + \frac{(\Omega_2)^2}{(\Omega_1 - \Omega_2) (\Omega_2 - \Omega_3)} + \frac{(\Omega_3)^2}{(\Omega_2 - \Omega_3) (\Omega_3 - \Omega_1)}
\end{equation}

In case of the Taub-NUT, we need to set $\Omega_2 = \Omega_3 = \Omega \neq \Omega_1 = \Lambda$. This way, we will get 
the following metric, system of equations and first integral:

\vspace{-0.5cm}
\begin{align}
d \widetilde{s}^2 &= \Omega^2 \Lambda \ d \widetilde{r}^2 + \Lambda \big[ \big( \sigma_2 \big)^2 + \big( \sigma_3 \big)^2 \big] + \frac{\Omega^2}{\Lambda} \big( \sigma_1 \big)^2 \\ \nonumber \\
\label{diffeq} & \frac{d \Lambda}{d \widetilde{r}} = \Omega ( \Omega - 2 \Lambda ) \hspace{2cm} \frac{d \Omega}{d \widetilde{r}} = - \Omega^2 \\ \nonumber \\
\bigg[ \lim_{\Omega_2 \rightarrow \Omega_3 = \Omega} Q \bigg]_{\Omega_1 = \Lambda} &= - \frac{\Lambda^2}{(\Lambda - \Omega)^2} + \frac1{\Lambda - \Omega} \bigg[ \lim_{\Omega_2 \rightarrow \Omega_3 = \Omega} \bigg( \frac{(\Omega_2)^2}{\Omega_2 - \Omega_3} - \frac{(\Omega_3)^2}{\Omega_2 - \Omega_3} \bigg) \bigg] \nonumber \\
&= - \frac{\Lambda^2}{(\Lambda - \Omega)^2} + \frac{2 \Omega}{\Lambda - \Omega} = - 1 - \bigg(\frac{\Omega}{\Lambda - \Omega} \bigg)^2
\end{align}

Rescaling the radius and solving (\ref{diffeq}) with suitable constants of integration gives us:

$$d \widetilde{r} = - \frac{d r}{2 m \Omega^2} \hspace{1.5cm} \frac{ d \Omega}{d r} = \frac1{2m} \hspace{1.5cm} \frac{d \ }{d r} \bigg( \frac{\Lambda}{\Omega^2} \bigg) = - \frac1{\Omega^2} \frac{ d \Omega}{d r}$$

\begin{equation}
\Omega = \frac{r - m}{2m} \hspace{2.5cm} \Lambda = \frac{r^2 - m^2}{4 m^2}
\end{equation}

and rescaling the metric as $d \widetilde{s} = \dfrac{d s}{2m}$ we get the Taub-NUT (\ref{tnuteq}) and conserved quantity:

$$d s^2 = \frac{r + m}{r - m} d r^2 + 4 m^2 \frac{r - m}{r + m} \big( \sigma_1 \big)^2+ (r^2 - m^2) \big[ \big( \sigma_2 \big)^2 + \big( \sigma_3 \big)^2 \big]$$

\begin{equation}
\bigg[ \lim_{\Omega_2 \rightarrow \Omega_3 = \Omega} Q \bigg]_{\Omega_1 = \Lambda} = - 1 - \bigg(\frac{\Omega}{\Lambda - \Omega} \bigg)^2 = - \frac{r^2 - 2 m r + 5 m^2}{(r - m)^2}
\end{equation}

This concludes another possible symmetry of the Taub-NUT as a member of Bianchi-IX 
metrics or solutions to Darboux-Halphen systems.

\subsection{Curvature and anti-self duality}

Now that we have identified the individual vierbeins, we shall proceed to compute the spin 
connections. We can describe the vierbeins as:

\begin{equation}
e^0 = c_0 (r) dr \hspace{1cm} e^i = c_i (r) \sigma^i, \hspace{0.5cm} i = 1, 2, 3
\end{equation}

Obviously, $e^0$ produces no connection terms $( d e^0 = 0 )$. Under torsion-free condition the 
1st Cartan structure equation $( d e^i = - {\omega^i}_j \wedge e^j )$ gives us the spin connections.

\begin{equation}
{\omega^i}_0 = \frac{\partial_r c_i}{c_0} \ \sigma^i \hspace{2cm} {\omega^i}_j = - {\varepsilon^i}_{jk} \frac{c_i^2 + c_j^2 - c_k^2 }{2 c_i c_j} \ \sigma^k
\end{equation}

The elaborate form of the spin connections is used to keep it anti-symmetric. We therefore 
construct the spin-connection matrix as shown below:

\vspace{-0.25cm}
\begin{equation}
\label{mat} \omega = 
\left({\begin{array}{cccc}
0 & - \frac{2m^2}{( r + m )^2} \sigma^1 & - \big( 1 - \frac m{r + m} \big) \sigma^2 & - \big( 1 - \frac m{r + m} \big) \sigma^3\\
\frac{2m^2}{( r + m )^2} \sigma^1 &  0 & - \frac m{r + m} \sigma^3 & \frac m{r + m} \sigma^2\\
\big( 1 - \frac m{r + m} \big) \sigma^2 & \frac m{r + m} \sigma^3 & 0 & - \big( 1 - \frac{2m^2}{(r + m)^2} \big) \sigma^1\\
\big( 1 - \frac m{r + m} \big) \sigma^3 & - \frac m{r + m} \sigma^2 & \big( 1 - \frac{2m^2}{(r + m)^2} \big) \sigma^1 & 0
\end{array} } \right)
\end{equation}

If we view the spin connections as a linear combination of self dual and anti-self dual tensors, 
then we can accordingly seperate out the self and anti-self dual components as $\omega_{ij} = \omega_{ij}^{(+)} + \omega_{ij}^{(-)}$. 
To this end, we can split the spin connection matrix (\ref{mat}) into two separate components: the 
self dual and the anti-self dual parts 

\vspace{-0.5cm}
\begin{equation} 
\label{spinupdown} 
\begin{split}
\omega^{(+)} &= - \frac12 \big( \sigma^1 \eta_1 + \sigma^2 \eta_2 + \sigma^3 \eta_3 \big) = - \frac12 \sigma^i \eta_i\\
\omega^{(-)} &= \bigg\{ \bigg( \frac12 - \frac{2m^2}{(r + m)^2} \bigg) \sigma^1 \bar{\eta}_1 - \bigg( \frac12 - \frac m{r + m} \bigg) \big( \sigma^2 \bar{\eta}_2 - \sigma^3 \bar{\eta}_3 \big) \bigg\}
\end{split}
\end{equation}

For reference, we have the t'Hooft symbol matrices $\eta^{(\pm)}$, which exhibit the $su(2)$ Lie algebra:

$$
\big[ \eta_i, \eta_j \big] = - 2 {\varepsilon_{ij}}^k \eta_k
$$

 The curvature tensor can be decomposed into self and anti-self dual parts $R_{ij} = R_{ij}^{(+)} + R_{ij}^{(-)}$, 
where according to Cartan's 2nd equation, $R = d \omega + \omega \wedge \omega$. Thus, we can write the spin 
connections as as a linear combination of self and anti-self dual t'Hooft symbols giving us the 
self-dual and anti-self dual spin connections described in (\ref{spinupdown}). Consequently, according to 
(\ref{struc}), the curvature tensor vanishes for the self-dual part:

\begin{equation}
\therefore \hspace{0.5cm} \label{sdcurv} R^{(+)} = d \omega^{(+)} + \omega^{(+)} \wedge \omega^{(+)} = - \frac12 \Big( d \sigma^i +  {\varepsilon^i}_{jk} \ \sigma^j \wedge \sigma^k \Big) \eta_i = 0
\end{equation}

Only the anti-self dual curvature remains, reflecting the Taub-NUT's anti-self dual nature. 
We make our job easier by writing the spin connection as $\omega^{(-)} = \omega^{(-)}_1 + \omega^{(-)}_2$:

\vspace{-0.35cm}
\begin{equation}
\omega^{(-)} = \frac12 \big( \sigma^1 \bar{\eta}_1 - \sigma^2 \bar{\eta}_2 + \sigma^3 \bar{\eta}_3 \big) + \bigg\{ - \frac{2m^2}{(r + m)^2} \sigma^1 \bar{\eta}_1 + \frac m{r + m} \big( \sigma^2 \bar{\eta}_2 - \sigma^3 \bar{\eta}_3 \big) \bigg\}
\end{equation}

where one can verify that $\omega^{(-)}_1$ will follow the same rule as $\omega^{(+)}$ in (\ref{sdcurv}). \\

 This allows us to compute the anti-self-dual curvature is given by

\vspace{-0.5cm}
\begin{equation}
\begin{split}
\label{riemt} \therefore \hspace{0.5cm} R^{(-)} &= \frac{2 m}{(r + m)^3} \bar{\eta}_1 \big( e^0 \wedge e^1 - e^2 \wedge e^3 \big)\\
& \hspace{1cm} + \frac m{(r + m)^3} \Big\{ - \bar{\eta}_2 \big( e^0 \wedge e^2 - e^3 \wedge e^1 \big) + \bar{\eta}_3 \big( e^0 \wedge e^3 - e^1 \wedge e^2 \big) \Big\}
\end{split}
\end{equation}

where we can see from the signs attached to the dual components that the curvature derived 
from Taub-NUT metric is clearly anti-self dual, as shown in \cite{sdgtn}. This also lets us conclude 
that it is an instanton. To elaborate further, we can show that only $SU(2)_-$ gauge fields are 
embedded within the spin-connection components as shown below:

\vspace{-0.5cm}
\begin{align}
\omega^{(\pm)}_{\mu \nu} = \eta^{(\pm)k}_{\mu \nu} A^{(\pm)}_k \hspace{1cm} \Rightarrow \hspace{1cm} &A^{(\pm)i} = \frac14 \eta^{(\pm)i}_{\mu \nu} \omega_{\mu \nu} \\ \nonumber \\ 
\begin{split}
A^{(+)1} = - \frac{\sigma^1}2 \hspace{2cm} A^{(-)1} &= \ \ \bigg( 1 - \frac{4m^2}{(r + m)^2} \bigg) \frac{\sigma^1}2 \\
A^{(+)2} = - \frac{\sigma^2}2 \hspace{2cm} A^{(-)2} &= - \frac{r - m}{r + m} \frac{\sigma^2}2 \\
A^{(+)3} = - \frac{\sigma^3}2 \hspace{2cm} A^{(-)3} &= \ \ \frac{r - m}{r + m} \frac{\sigma^3}2
\end{split}
\end{align}

while the field strengths  are given by:

\begin{equation}
R^{(-)}_{\mu \nu} = \eta^{(-)k}_{\mu \nu} F^{(-)}_k \hspace{1cm} \Rightarrow \hspace{1cm} F^{(\pm)i} = \frac14 \eta^{(\pm)i}_{\mu \nu} R_{\mu \nu}
\end{equation}

\vspace{-0.5cm}
\begin{equation}
\begin{split}
F^{(-)1} = R_{01} &= - R_{23} = \ \ \frac{2m}{(r + m)^3} \big( e^0 \wedge e^1 - e^2 \wedge e^3 \big) \\
F^{(-)2} = R_{02} &= - R_{31} = - \frac{m}{(r + m)^3} \big( e^0 \wedge e^2 - e^3 \wedge e^1 \big) \\
F^{(-)3} = R_{03} &= - R_{12} = \ \ \frac{m}{(r + m)^3} \big( e^0 \wedge e^3 - e^1 \wedge e^2 \big) 
\end{split}
\end{equation}

where it is obvious that due to the absence of self-dual curvature, there are no $SU(2)_+$ 
gauge fields, ie. $F^{(+)i} = 0$ and thus field strengths are anti-self dual ($F= -*F$) which 
of course, coincide with the curvature tensor (\ref{riemt}). In terms of 2-forms, the independent 
components are given by :

\vspace{-0.5cm}
\begin{align}
R^{(-)}_{0101} &= R^{(-)}_{2323} = - R^{(-)}_{0123} = \frac{2 m}{(r + m)^3}\\
R^{(-)}_{0202} &= R^{(-)}_{1313} = R^{(-)}_{0213} = - \frac m{(r + m)^3}\\
R^{(-)}_{0303} &= R^{(-)}_{1212} = - R^{(-)}_{0213} = - \frac m{(r + m)^3}
\end{align}

This lets us compute the Ricci tensors and scalar in accordance with the formula:

\vspace{-0.5cm}
\begin{align}
\mathcal{R}_{ik} = g^{jl} \mathcal{R}_{ijkl} &= \delta^{jl} \mathcal{R}_{ijkl} \hspace{1cm} R = \delta^{ik} \mathcal{R}_{ik} \\  \nonumber \\
\therefore \hspace{0.5cm} \mathcal{R}_{00} = \mathcal{R}_{11} = \mathcal{R}_{22} &= \mathcal{R}_{33} = 0 \hspace{2cm} R = 0
\end{align}

Since  Ricci tensors vanish, Taub-NUT  is clearly a vaccum solution of Einstein's equations.

\subsection{Topological Invariants}

Topological invariants are analogous to an overall charge distributed in the manifold. In the 
gravity side, there are two topological invariants associated with the Atiyah-Patodi-Singer index 
theorem for a four dimensional elliptic complex \cite{egh-report,besse}: the Euler characteristic $\chi(M)$ and 
the Hirzebruch signature $\tau(M)$, which can be expressed as integrals of  four-manifold curvature. \\

Recall that in electromagnetic theory, the field action is given by:

\vspace{-0.5cm}
\[ \begin{split}
S &= - \frac1{16 \pi} \int d \Omega \ F_{ij} F^{ij} = - \frac1{16 \pi} \int F \wedge F \\ 
\text{where} \hspace{1cm} F = \frac{1}{2} F_{ij} dx^i &\wedge dx^j \hspace{1cm} \text{and} \hspace{1cm} \varepsilon^{ijkl} d \Omega = dx^i \wedge dx^j \wedge dx^k \wedge dx^l 
\end{split} \]

The equations of motion are obtained by solving for minimum variation of the electromagnetic 
field action. We merely apply these equations to compute topological invariants as 
integrals analogous to action. We can write for the general lagrangian:

\vspace{-0.25cm}
\begin{align}
\mathcal{L} &= c^{abcd} R_{ab} \wedge R_{cd} = c^{abcd} {F^{(\pm)m}_{ab}} {F^{(\pm)n}_{cd}} \eta^{(\pm)m}_{ij} \eta^{(\pm)n}_{kl} \varepsilon^{ijkl} d \Omega \nonumber \\ 
&= \pm 2 d \Omega c^{abcd} {F^{(\pm)m}_{ab}} \partial_c {A^{(\pm)m}_d} \hspace{1.5cm} \big( \text{where} \hspace{0.5cm} \varepsilon^{ijkl} d \Omega  = e^i \wedge e^j \wedge e^k \wedge e^l \big) 
\end{align}

Applying Lagrange's equation gives the contracted Bianchi identity for curvature as:

\vspace{-0.25cm}
\begin{equation}
\label{conb} \partial_c \bigg( \frac{\partial \mathcal{L}}{\partial (\partial_c A^{(\pm)m}_d)} \bigg) = \pm 2 c^{abcd} \partial_c {F^{(\pm)m}_{ab}} = 0
\end{equation}

Conversely, we can say that Bianchi identity for $SU(2)_\pm$ gauge fields is at the root of the 
invariance of topological quantities. One can verify this starting from (\ref{conb}) and then working 
in reverse order to obtain the invariants. 

 Given that the boundary integral vanishes, the overall invariant is computed  only from the 
bulk part. For non-compact manifolds like Taub-NUT, there are additional boundary terms 
neither separated into self-dual nor anti-self-dual parts unlike the volume terms. They are the 
so-called  eta-invariant $\eta_S(\partial M)$, given for $k$ self-dual gravitational instantons by \cite{gibb-perry}

\vspace{-0.35cm}
\begin{equation}\label{eta-inv}
    \eta_S(\partial M) = - \frac{2 \epsilon}{3k} + \frac{(k-1)(k-2)}{3k} \hspace{0.5cm} 
\begin{cases}
\epsilon =0 ; \hspace{0.5cm} \text{ALE boundary conditions} \\
\epsilon = 1 ; \hspace{0.5cm} \text{ALF boundary conditions}
\end{cases}
\end{equation}

Since Taub-NUT is a ALF hyper-kahler four-manifold it has a non-vanishing eta-invariant 
which is equal to $-\frac{2}{3}$. According to calculations described in \cite{ymgi}, upon applying curvature 
components of (\ref{riemt}), the Euler characteristic $\chi$ and the Hirzebruch signature complex $\tau$ are:

\vspace{-0.5cm}
\begin{align}
\chi (M) &= \frac1{32 \pi^2} \int_M \varepsilon^{abcd} R_{ab} \wedge R_{cd}  = 1 \\ \nonumber \\
\begin{split}
\tau_{bulk} (M) &= - \frac1{12 \pi^2} \bigg( \int_M R_{ab} \wedge R_{ab} \bigg)_{a < b} = \frac23 \\ 
\therefore \hspace{1cm} &\tau(M) = \tau_{bulk} (M) + \eta_S (\partial M) = 0
\end{split}
\end{align} 

One could say that the general form of various topological invariants can be written as:

\vspace{-0.5cm}
\begin{equation}
\mathcal{C} ( M ) = \frac1{k \pi^2} \int_M c^{abcd} R_{ab} \wedge R_{cd} =
\begin{cases}
\dfrac1{k \pi^2} \int_M F_{ab} \big(* F^{ab} \big); \hspace{0.35cm} c^{abcd} = \varepsilon^{abcd} \hspace{0.25cm} \text{\big(Euler Char.\big)}\\
\dfrac1{k \pi^2} \int_M  F_{ab} F^{ab}; \hspace{0.85cm} c^{abcd} = g^{ac} g^{bd} \hspace{0.25cm} \text{\big(Hirzebruch Sign.\big)}
\end{cases}
\end{equation} 

where $c^{abcd}$ is contracting tensor defined in respect to the relevant circumstances.

\numberwithin{equation}{section}

\section{Killing-Yano tensors and the Taub-NUT metric}

There are tensors quadratic in momenta and conserved along geodesics, expressed as a vector 
$\bm{K}$ whose components transform among themselves under 3-dimensional rotations. They are 
very similar to the Runge-Lenz vector in the Kepler problem with components:

\vspace{-0.25cm}
\begin{equation}
\label{kyano} K^{(i)} = \frac12 K^{(i) \mu \nu} p_\mu p_\nu
\end{equation}

Provided that $J^0 \neq 0$, such vectors usually satisfy the following property:

\vspace{-0.25cm}
\begin{equation}
\label{conserved} \bm{r} . \bigg( \bm{K} \pm \frac{H \bm{J}}{J^{0}} \bigg) = \frac12 \Big( \bm{J}^2 - \big( J^{0} \big)^2 \Big)
\end{equation}

where if $(J^0, \bm{J}, H, \bm{K})$ are all constant, the 3-dimensional position vector $\bm{r}$ lies in a plane. 
Using (\ref{conserved}) and the relation $J^0 = \dfrac{\bm{r}.\bm{J}}r$, we can see that:

\vspace{-0.25cm}
\begin{equation}
\label{conserv} \bm{r} . \bm{K} = \mp r H + \frac12 \Big( \bm{J}^2 - \big( J^{0} \big)^2 \Big)
\end{equation}

In Taub-NUT geometry, there are also 4 completely antisymmetric Killing tensors known 
as Killing-Yano (KY) tensors. Three of these are complex structures, realizing quaternionic 
algebra since the Taub-NUT manifold is hyper-K$\ddot{\text{a}}$hler. The fourth is a scalar with a non-
vanishing field strength and it exists by virtue of the metric being of Petrov type D. Their 
existence is implied by a triplet of symmetric 2nd rank Killing tensors called the St$\ddot{\text{a}}$ckel-Killing 
tensor satisfying:

\vspace{-0.25cm}
\begin{equation}
\label{kyano} D_{( \lambda} K^{(i)}_{\mu \nu )} = 0
\end{equation}

We will examine properties of KY tensors relevant for studying Taub-NUT symmetries. 
Before that, let us list some references that initiated the study of such dynamical symmetries.

Dynamical symmetries of the Kaluza Klein monopole were discussed in detail by Feher 
in \cite{Feher_dynamical}. The dynamics of two non-relativistic BPS monopoles was described using Atiyah-
Hitchin metric (Taub-NUT being a special case), the corresponding $O(4)/ O(3,1)$ symmetry 
discovered in \cite{Gibbons:1986df}, and applied to calculate the underlined motion group-theoretically in \cite{Feher:1986ib}. 
The symmetry was then extended to $O(4,2)$ in \cite{Gibbons:1987sp} and \cite{Cordani:1988ps}. In \cite{Gibbons:1987sp} Gibbons et. al discussed 
dynamical symmetries of multi-centre metrics and applied the results to the scattering of BPS 
monopoles and fluctuations around them, giving a detailed account of the hidden symmetries of 
the Taub-NUT. The hidden symmetries in large-distance interactions between BPS monopoles 
and of the fluctuations around them are traced to the existence of a KY tensor on the self-dual 
Taub-NUT. The global action on classical phase space of these symmetries was discussed in 
\cite{Gibbons_PLB87} and the quantum picture involving the "dynamical groups" $SO(4)$, $SO(4, 1 )$ and $SO(4, 2)$ 
was also given. A comprehensive review of the dynamical symmetry can be found  in \cite{Horvathy_review}. 
Supersymmetry and extension to spin has also been studied in \cite{Comtet:1995zd, Holten_susy}.

\numberwithin{equation}{subsection}

\subsection{Yano and St$\ddot{\text{a}}$ckel tensors}

We can construct these KY tensors in terms of simpler objects known as Yano tensors that 
are antisymmetric rank 2 tensors satisfying the Killing like equation. Thus, the covariant 
derivative is antisymmetric over permutations of all possible pairs of indices. This allows us 
to write the covariant derivative of the Yano tensor in terms of the cyclic permutations as:

\vspace{-0.5cm}
\begin{align} \displaybreak[0]
\label{yano1} f_{\mu \nu} &= - f_{\nu \mu} \hspace{2cm} \nabla_\mu f_{\nu \lambda} + \nabla_\nu f_{\mu \lambda} = 0 \\ 
\label{covd1} \nabla_\mu f_{\nu \lambda} = \nabla_\nu f_{\lambda \mu} &= \nabla_\lambda f_{\mu \nu} = \nabla_{[ \mu} f_{\nu \lambda ]} \hspace{0.5cm} = \hspace{0.5cm} \frac13 \big( \nabla_\mu f_{\nu \lambda} + \nabla_\nu f_{\lambda \mu} + \nabla_\lambda f_{\mu \nu} \big)
\end{align}

We can also construct symmetric Killing tensors of rank 2 by symmetrized multiplication:

\begin{equation}
\label{syano} K^{(ab)}_{\mu \nu} = \frac12 \big( f^{(a)\lambda}_\mu f^{(b)}_{\lambda \nu} + f^{(b)\lambda}_\mu f^{(a)}_{\lambda \nu} \big) \hspace{0.25cm} \equiv \hspace{0.25cm} \frac12 \big( f^{(a)\lambda}_\mu f^{(b)}_{\lambda \nu} + f^{(a)\lambda}_\nu f^{(b)}_{\lambda \mu} \big) = K^{ab}_{(\mu \nu)}
\end{equation}

These symmetric Killing tensors satisfy the KY condition (\ref{kyano}). The Taub-NUT manifold 
admits 4 such KY tensors, given by a scalar $f^0$ and three components that transform as a 
vector $f^i$ $\forall \ i = 1, 2, 3$. We can form triplets of symmetric Killing tensors as in (\ref{syano}), given 
by setting $a = 0$ and $b = i$:

\vspace{-0.25cm}
\begin{equation}
\label{symk} K^{(i)}_{\mu \nu} = K^{(0i)}_{\mu \nu} = \frac12 \big( f^{0\lambda}_\mu f^{i}_{\lambda \nu} + f^{i\lambda}_\mu f^{0}_{\lambda \nu} \big) \hspace{1cm} i = 1, 2, 3
\end{equation}

Using (\ref{yano1}) we can see how they obey (\ref{kyano}) as follows:

\vspace{-0.2cm}
\[ \nabla_\gamma K^{ij}_{(\mu \nu)} + \nabla_\mu K^{ij}_{(\nu \gamma)} + \nabla_\nu K^{ij}_{(\gamma \mu)} = 0 \]

\begin{equation}
\nabla_{(\gamma} K^{ij}_{\mu \nu)} = 0 \hspace{0.5cm} \Rightarrow \hspace{0.5cm} \nabla_{(\gamma} K^i_{\mu \nu)} \equiv \nabla_{(\gamma} K^{0i}_{\mu \nu)} = 0
\end{equation}

Thus, we can feel assured that (\ref{kyano}) is satisfied by this symmetric Killing tensor. This 
allows us to construct the tensors of (\ref{kyano}) that are quadratic in momenta, showing how to 
get St$\ddot{\text{a}}$ckel tensors from the KY tensors. However, since the KY tensor is anti-symmetric, it 
cannot be used to form polynomials with components of the same vector. Thus, it will have to 
be a mixed product of components of different vector quantities, as found in case of the angular 
momentum, a product between one position and one momentum component each. Applying 
Holten's algorithm yields the Killing equation in (\ref{yano1}).

\subsection{Euclidean Taub-NUT}

The Taub-NUT metric \cite{tnky} admits four such Yano tensors written as the following 2-forms:

\vspace{-0.5cm}
\begin{align}
f^0 &= 4 \big( d \psi + \cos \theta \ d \phi \big) \wedge dr + 2r \big( r \pm 1 \big) \big( r \pm 2 \big) \sin \theta \ d \theta \wedge d \phi \\
\label{tnky} f^i &= \pm 4 \big( d \psi + \cos \theta \ d \phi \big) \wedge d x^i - {\varepsilon^i}_{jk} f(r) \ dx^j \wedge dx^k, \hspace{1cm} \forall i, j, k = 1, 2, 3
\end{align}

One can always find Killing tensors embedded within conserved quantities, as evident from 
the Poisson Brackets of any conserved quantity expanded ala Holten algorithm. The coefficient 
from Laplace-Runge-Lenz vector is analogous to the Killing-St$\ddot{\text{a}}$ckel tensor $K_{ij}$, so we can argue:

\begin{equation}
Q^{(2)} = K_{ij} \Pi^i \Pi^j \equiv \frac12 C^{(2)}_{ij} \Pi^i \Pi^j
\end{equation}

Now the angular momentum co-efficients according to (\ref{amom}) are:

\begin{equation}
C^{(0)} = q \ g_{jk} (\vec{x}) \ \frac{x^j}r \theta^k  \hspace{1cm} C^{(1)}_i = - g_{im} (\vec{x}) {\varepsilon^m}_{jk} \theta^k x^j
\end{equation}

If we write $C^{(1)}_i = f_{ik} \theta^k$  (see section \textbf{8.1} in APPENDIX), using Holten's Algorithm gives:

\vspace{-0.5cm}
\[ \begin{split}
\nabla_j C^{(1)}_i &= \nabla_j f_{ik} \theta^k = - g_{im} (\vec{x}) {\varepsilon^m}_{jk} \theta^k \\ 
\nabla_i C^{(1)}_j + \nabla_j C^{(1)}_i &= 0 \hspace{1cm} \Rightarrow \hspace{1cm} \big( \nabla_i f_{jk} + \nabla_j f_{ik} \big) \theta^k = 0
\end{split} \]

which is the Killing equation (\ref{yano1}). Thus, we can say that the KY tensor is

\vspace{-0.5cm}
\begin{align}
f^0_{jk} &= g_{jk} ( \vec{x} ) \hspace{1cm} \Rightarrow \hspace{1cm} f^j_{0k} = \delta^j_k \\
f^i_{jk} &= {\varepsilon^i}_{jk} \hspace{1cm} \Rightarrow \hspace{1cm} f^i = {\varepsilon^i}_{jk} e^j \wedge e^k
\end{align}

such that the square of it gives the St$\ddot{\text{a}}$ckel tensor

\begin{equation}
K_{ij}^{k} = {f^0}_{im} {f^{km}}_j
\end{equation}

This shows how Killing tensors are embedded within the conserved quantities. We can 
choose four combinations of three indices out of the available four. Since Taub-NUT can be 
written in an alternate form given by (\ref{tn}), the vierbeins of the metric are given by:

\vspace{-0.2cm}
\begin{equation}
\label{vbns} e^0 = \frac{4 \big( d \psi + \vec{A} . d \vec{x} \big)}{\sqrt{f(r)}} \hspace{1cm} e^i = \sqrt{f(r)} \ dx^i
\end{equation} 

So, according to our theory, we should have

\vspace{-0.5cm}
\begin{equation}
\begin{split}
f^{i} &= - {\varepsilon^i}_{jk} \ e^j \wedge e^k + {\delta^i}_k \ e^0 \wedge e^k \\
&= - {\varepsilon^i}_{jk} f(r) dx^j \wedge dx^k \pm  4 \big( d \psi + \vec{A} . d \vec{x} \big) \wedge d x^i
\end{split}
\end{equation}

This result so far is comparable with the result (\ref{tnky}), so we have a possible method for 
constructing Killing-Yano tensors from the coefficients of conserved quantities. Their covariant 
exterior derivatives and their properties are given by:

\vspace{-0.5cm}
\begin{align}
D f^0 &= \nabla_\gamma f^0_{\mu \nu} dx^\gamma \wedge dx^\mu \wedge dx^\nu = r \big( r \pm 2 \big) \sin \theta dr \wedge d \theta \wedge d \phi\\
D f^i &= 0, \hspace{1.5cm} \forall i = 1, 2, 3
\end{align}

From the results above, we can infer that the covariant derivatives hold following properties:

\vspace{-0.2cm}
\begin{equation}
\nabla_\gamma f^0_{\mu \nu} = \nabla_\mu f^0_{\nu \gamma} = \nabla_\gamma \nu f^0_{\gamma \mu} \hspace{1.5cm} \nabla_\gamma f^i_{\mu \nu} = 0 \hspace{1cm} i = 1, 2, 3
\end{equation}

showing that they obey the condition for covariant derivatives of KY tensors. As shown 
in (\ref{symk}), these tensors can form a symmetric triplet or a vector of Killing tensors. They also 
exhibit the mutual anti-commutation property:

\vspace{-0.2cm}
\begin{equation}
\label{quatern} f^i f^j = - \delta^{ij} + {\varepsilon^{ij}}_k f^k \hspace{1cm}
\begin{cases}
\big\{ f^i, f^j \big\} = f^i f^j + f^j f^i = - 2 \delta^{ij} \\
\big[ f^i, f^j \big] = f^i f^j - f^j f^i = 2 {\varepsilon^{ij}}_k f^k
\end{cases}
\end{equation}

proving that they are complex structures realizing the quaternion algebra. This implies that 
the 2-forms $f^i$ are objects in the quaternionic geometry and possibly \textbf{hyperk$\ddot{\text{a}}$hler} structures. 
This leads us to the next section where we examine the hyperk$\ddot{\text{a}}$hler structure of the Taub-NUT.

\subsection{Graded Lie-algebra via Schouten-Nijenhuis Brackets}

We will now see if the KY tensors of the Taub-NUT metric exhibit Lie algebra under the action of Schouten-Nijenhuis Brackets. If they do, it would allow us to form higher order KY tensors from lower order ones of rank greater than 1. In particular it is noteworthy in this context that, Kastor et. al already found that KY tensors on constant curvature spacetimes do form Lie algebras with respect to the SN bracket \cite{Kastor:2007tg}.  \\

The Schouten-Nijenhuis Bracket (SNB) is a bracket operation between multivector fields. The SNB for two such fields $A = A^{i_1 i_2 . . . i_m} \bigwedge_{k=1}^m \partial_{i_k} \ ; \ B = B^{j_1 j_2 . . . j_n} \bigwedge_{k=1}^n \partial_{j_k}$, is given by

\vspace{-0.5cm}
\begin{equation}
\begin{split}
C^{a_1 . . . a_{m+n-1}} &= \big[ A, B \big]^{a_1 . . . a_{m+n-1}}_{SN}\\
= m &A^{c [ a_1 . . . a_{m - 1}} \nabla_c B^{a_m . . . a_{m + n - 1}]} + n \big( -1 \big)^{mn} B^{c [ a_1 . . . a_{n - 1}} \nabla_c A^{a_n . . . a_{m + n - 1}]}
\end{split}
\end{equation}

This new tensor is completely antisymmetric, fulfilling the first requirement to be considered a KY tensor. All that remains is for its covariant derivative to exhibit the same Killing equation (\ref{covd1}) relevant to such tensors. Now, we will use an important identity (see (\ref{dderiv}) in APPENDIX) for KY tensors:

\vspace{-0.25cm}
\begin{equation}
\therefore \hspace{1cm} \nabla_a \nabla_b K_{c_1 c_2 . . . c_n} = ( - 1 )^{n + 1} \frac{n+1}2 {R_{[bc_1|a|}}^d K_{c_2 c_3 . . . c_n ] d}
\end{equation}

we get upon applying to the covariant derivative of this new tensor:

\vspace{-0.5cm}
\begin{equation}
\begin{split}
\label{kysn} \nabla_b C_{a_1 . . . a_{m+n-1}} = - \big( m + n \big) \big( \nabla_c A_{[ b a_1 . . . a_{m-1}} \big) &\nabla^c B_{a_m . . . a_{m+n-1}]} \\
 - \big( m + n \big) &{A^c}_{[a_1 . . . a_{m-1}} R_{|bd|c a_m} {B_{a_{m+1} . . . a_{m+n-1}]}}^d
\end{split}
\end{equation}

The 1st term easily shows anti-symmetry of index $b$ with other indices, but the 2nd term exhibits it only under certain circumstances. One could say that by symmetry properties of the curvature tensor, in maximally symmetric spaces it could be expressed as:

\vspace{-0.25cm}
\begin{equation}
R_{abcd} (x) = f(x) g_{ij} (x) {\varepsilon^i}_{ab} {\varepsilon^j}_{cd} = f(x) \big\{ g_{ac} (x) g_{bd} (x) - g_{ad} (x) g_{bc} (x) \big\}
\end{equation}

So, for cases of constant curvature $f(x) = k$, we could write

\vspace{-0.2cm}
\begin{equation}
\label{ccurv} \big( R_{abcd} \big)_{const} = k \big\{ g_{ac} (x) g_{bd} (x) - g_{ad} (x) g_{bc} (x) \big\}
\end{equation}

Thus, upon applying the constant curvature formula of (\ref{ccurv}) to (\ref{kysn}), we will get

\vspace{-0.5cm}
\begin{equation}
\begin{split}
\nabla_b C_{a_1 . . . a_{m+n-1}} = - \big( m + n \big) &\big[ \big( \nabla_c A_{[ b a_1 . . . a_{m-1}} \big) \nabla^c B_{a_m . . . a_{m+n-1}]} \\
 &- k A_{[a_1 . . . a_{m-1}} {B_{a_m . . . a_{m+n-1} b]}} \big] = \nabla_{[b} C_{a_1 . . . a_{m+n-1}]}
\end{split}
\end{equation}

Clearly this matches the property eq.(\ref{covd1}) expresses, showing that it is also a KY tensor. So the SNB of any two KY tensors in spaces of constant curvature is also a KY tensor. 

However, as evident from (\ref{riemt}), the curvature of the Taub-NUT metric is not constant, allowing us to conclude that its KY tensors do not exhibit Lie algebra under SN Brackets. Thus, we cannot produce higher order KY tensors using the lower order ones for the Taub-NUT as shown in \cite{kyapp}. So, we are limited to the set of four available rank two  KY tensors.

\numberwithin{equation}{section}

\section{Hyperk\"ahler structure and the KY tensors}

Now we will consider the hyperk\"ahler structures related to the Taub-NUT metric. To begin with, we will define both, k$\ddot{\text{a}}$hler and hyperk\"ahler structures.\\

{\raggedleft\textbf{K$\ddot{\text{a}}$hler manifold:} If a complex manifold $M$ has a hermitian metric $g$ and a fundamental 2-form $\omega$ which is closed ($d \omega = 0$) then $M$ is a K$\ddot{\text{a}}$hler manifold and $\omega$ is a K$\ddot{\text{a}}$hler form.}\\

The connection between the metric $g$ and the K\"ahler form $\omega$ is:

\vspace{-0.25cm}
\begin{equation}
\omega_{\mu \nu} = {J_\mu}^\lambda . g_{\lambda \nu} = \big( J g \big)_{\mu \nu}
\end{equation}

where $J$ is the complex structure, for which $J^2 = -1$.\\

{\raggedleft\textbf{Hyperk\"ahler manifold:}} If $M$ is a hyper-complex manifold with a hyper-Hermitian metric g and a triplet of fundamental forms $\vec{\omega}$ which are closed ($d\vec{ \omega} = 0$)  then $M$ is a Hyperk$\ddot{\text{a}}$hler manifold. It is the same as the K\"ahler manifold except that there are more than one type of complex structures. In case of 4 dimensions, there are 3 such integrable complex structures ($i, j, k$), and they obey the algebraic relations:

\vspace{-0.125cm}
\begin{equation}
i^2 = j^2 = k^2 = ijk = -1
\end{equation}

This would also imply that there are corresponding number of different 2-forms available in this case, known as the Hyperk$\ddot{\text{a}}$hler forms, given by:

\begin{equation}
\label{hkah} \omega^i_{\mu \nu} = {J^i_\mu}^\lambda . g_{\lambda \nu} = \big( J^i g \big)_{\mu \nu}
\end{equation}

where $g_{\lambda \nu}$ is the hyper-hermitian metric and $J^i_{\mu \lambda}$ is the almost complex structure exhibiting quaternion algebra 

\vspace{-0.25cm}
\begin{equation}
J_\alpha J_\beta = - \delta_{\alpha \beta} I + {\varepsilon_{\alpha \beta}}^\gamma J_\gamma
\end{equation}

and thus, we can see that the hyperk$\ddot{\text{a}}$hler structures exhibit the same algebra:

\vspace{-0.5cm}
\[ \begin{split}
\big( J^i J^j \big)_{\mu \nu} &= J^i_{\mu \rho} \ g^{\rho \sigma} J^j_{\sigma \nu} \hspace{1cm} \big[ J^i, J^j \big]_{\mu \nu} = 2 {\varepsilon^{ij}}_k J^k_{\mu \nu} \\ 
\Big( \omega^i \omega^j \Big)_{\mu \nu} = \omega^i_{\mu \gamma} {\omega^{j \gamma}}_\nu &= \Big( {J^i_\mu}^\rho . g_{\rho \gamma} \Big) g^{\gamma \lambda} \Big( {J^j_\lambda}^\sigma . g_{\sigma \nu} \Big) = {J^i_\mu}^\rho {J^j_\rho}^\sigma g_{\sigma \nu} = \big( J^i J^j g \big)_{\mu \nu} \\ 
\therefore \hspace{1cm} \big[ \omega^i, \omega^j \big]_{\mu \nu} &= \Big( \big[ J^i, J^j \big] g \Big)_{\mu \nu} = 2 \big( {\varepsilon^{ij}}_k J^k g \big)_{\mu \nu} = 2 {\varepsilon^{ij}}_k \omega^k_{\mu \nu}
\end{split} \]

These complex structures originate from the t'Hooft symbols which have 3 self dual and 3 anti-self dual components. That means we could have six different symplectic 2-forms. The almost complex structures $J^i$ can be represented by t'Hooft symbols, which themselves can be given by linear combinations of antisymmetric tensor ${\varepsilon^i}_{jk}$ and delta function ${\delta^i}_j$. 

\vspace{-0.25cm}
\begin{equation}
J^i_{jk} = {\varepsilon^i}_{jk} \pm \frac12 \big( {\delta^0}_j {\delta^i}_k - {\delta^0}_k {\delta^i}_j \big)
\end{equation}

Thus, we can argue that hyper-k$\ddot{\text{a}}$hler structures given by (\ref{hkah}) are:

\vspace{-0.25cm}
\begin{equation}
\omega^i_{jk} = \big( J^i g \big)_{jk} = g_{jn} (\vec{x}) \bigg[ {\varepsilon^{in}}_k  \pm \frac12 \big( \delta^{0n} {\delta^i}_k - {\delta^0}_k {\delta^{in}} \big) \bigg]
\end{equation}

As introduced in (\ref{altn}) and following \cite{hkstr} we shall take a different form of the Taub-NUT

\vspace{-0.25cm}
\begin{equation}
ds^2 = V(r) \ \delta_{ij} \ dx^i dx^j + V^{-1} (r) \big( d \tau + \vec{\sigma} . d \vec{r} \big)^2
\end{equation}

for which, the vierbeins, in a similar fashion to (\ref{vbns}) are given by

\vspace{-0.25cm}
\[ e^0 = \frac{4 \big( d \tau + \vec{\sigma} . d \vec{r} \big)}{\sqrt{V(r)}} \hspace{1cm} e^i = \sqrt{V(r)} \ dx^i \]

Thus, remembering that $g = \delta_{ij} e^i \otimes e^j$ the hyper-k$\ddot{\text{a}}$hler forms are given by:

\vspace{-0.25cm}
\begin{equation}
\omega^i = \omega^i_{jk} dx^j \wedge dx^k = J^i_{jk} e^j \wedge e^k
\end{equation}

\vspace{-0.125cm}
\[ \omega^i = \bigg[ {\varepsilon^i}_{jk} \pm \frac12 \big( {\delta^0}_j {\delta^i}_k - {\delta^0}_k {\delta^i}_j \big) \bigg] e^j \wedge e^k \hspace{0.25cm} = \hspace{0.25cm} {\varepsilon^i}_{jk} V(r) dx^j \wedge dx^k - e^0 \wedge e^i \]

\vspace{-0.125cm}
\begin{equation}
\therefore \hspace{1cm} \omega^i = {\varepsilon^i}_{jk} V(r) dx^j \wedge dx^k \pm \big( d \tau \wedge dx^i + \sigma_n . d x^n \wedge dx^i \big)
\end{equation}

For the Taub-NUT, choosing only anti-self-dual components for $V(r) = l + \dfrac1r$ and restricting $\vec{\sigma}$ to lie on a plane ($\vec{\sigma} = ( 0, \sigma_2, \sigma_3 )$), the reduced symplectic forms are:

\vspace{-0.5cm}
\begin{equation}
\begin{split}
\omega^1 &= dx^1 \wedge d \tau + \sigma_2 dx^1 \wedge dx^2 + \sigma_3 dx^1 \wedge dx^3 + \bigg( l + \frac1r \bigg) dx^2 \wedge dx^3 \\
\omega^2 &= dx^2 \wedge d \tau + \sigma_3 dx^2 \wedge dx^3 - \bigg( l + \frac1r \bigg) dx^1 \wedge dx^3 \\
\omega^3 &= dx^3 \wedge d \tau - \sigma_2 dx^2 \wedge dx^3 + \bigg( l + \frac1r \bigg) dx^1 \wedge dx^2
\end{split}
\end{equation}

This construction of  hyperk\"ahler structures is similar to how spatial KY tensors were deduced, proving that the KY tensors are the hyperk$\ddot{\text{a}}$hler structures of the Taub-NUT metric.

\begin{center}
\textbf{Comparison between Killing-Yano Tensors and Hyperk$\ddot{\text{a}}$hler Structures}\\

\begin{tabular}{|c|c|c|}

\hline & & \\

$\bm{i}$ & \textbf{Killing-Yano tensor} $\bm{f^i}$ & \textbf{Hyperk$\ddot{\text{a}}$hler structure} $\bm{\omega^i}$ \\
 & & \\

\hline & & \\

$i$
&
$\pm 4 \big( d \psi + A_n d x^n  \big) \wedge d x^i - {\varepsilon^i}_{jk} f(r) dx^j \wedge dx^k$
&
$\pm \big( d \tau + \sigma_n . d x^n \big) \wedge dx^i + {\varepsilon^i}_{jk} V(r) dx^j \wedge dx^k $\\
& & \\

$1$
&
$\mp 4 dx^1 \wedge \big( d \psi + A_n dx^n \big) + \bigg( 1 + \dfrac4r\bigg) dx^2 \wedge dx^3$
&
$dx^1 \wedge \big( d \tau + \sigma_2 dx^2 + \sigma_3 dx^3 \big) + \bigg( l + \dfrac1r \bigg) dx^2 \wedge dx^3 $\\

& & \\

$2$
&
$\mp 4 dx^2 \wedge \big( d \psi + A_n dx^n \big) - \bigg( 1 + \dfrac4r \bigg) dx^1 \wedge dx^3$
&
$dx^2 \wedge \big( d \tau + \sigma_3 dx^3 \big) - \bigg( l + \dfrac1r \bigg) dx^1 \wedge dx^3$\\
& & \\

$3$
&
$\mp 4 dx^3 \wedge \big( d \psi + A_n dx^n \big) + \bigg( 1 + \dfrac4r \bigg) dx^1 \wedge dx^2$
&
$dx^3 \wedge \big( d \tau - \sigma_2 dx^3 \big) + \bigg( l + \dfrac1r \bigg) dx^1 \wedge dx^2$\\
& & \\

\hline
\end{tabular} 
\end{center}

Few points are worth mentioning here. By studying the $G_2$ holonomy equation for biaxial anti-self dual Bianchi IX base Gibbons et.al \cite{CGLP} found that the associated first order 
equations satisfied by the metric coefficients yield the self-dual Ricci flat Taub-NUT metrics where $SO(3) \subset  U(2)$ rotates the three hyperk$\ddot{\text{a}}$hler forms as a triplet.

\section{Discussion}

In this article we see that the Taub-NUT is comparable to Euclideanized Bertrand spacetime with magnetic fields due to the shared geometry and conserved quantities, and a dual configuration as either Oscillator or Kepler systems.
Identical conserved quantities are connected to identical symmetries and Killing tensors embedded within. These are the Killing-St$\ddot{\text{a}}$ckel and Killing-Yano tensors embedded as co-efficients within the Laplace-Runge-Lenz and angular momentum vectors respectively. The Killing-Yano tensors exhibit quaternionic algebra, hinting at a link between them and hyperk$\ddot{\text{a}}$hler structures matching the form of the KY tensors derived from the angular momentum. This confirms that the KY tensor and hyperk$\ddot{\text{a}}$hler structures are the same for Taub-NUT. Since symmetries of a spacetime are unaffected upon euclideanization, we can expect that all properties arising from shared symmetries are also exhibited by Bertrand spacetimes with magnetic fields.

Taub-NUT is a special case of the anti-self-dual Bianchi-IX spaces \cite{cgr1}, derived by solving the dynamical equations that emerge from applying the settings for this case to the related classical Darboux-Halphen system. The shared geometric properties, including the Ricci flow, integrability aspects and integrable reductions to Painleve systems can be explored to some extent. In special situations, self-dual Einstein Bianchi-IX metrics reduce to Taub NUT de Sitter metric with two parameters of the biaxial solutions respectively identified as the NUT parameter and the cosmological constant. Computing its curvature confirms that the Taub-NUT is anti-self dual, reflecting its instantonic nature, and expectedly, Ricci-flat with topological invariants to compare with other possible diffeomorphically equivalent Ricci-flat manifolds. According to Kronheimer classifications \cite{kron1, kron2} all  hyperk$\ddot{\text{a}}$hler metrics like Taub-NUT in four dimensions are always anti-self dual, so the  hyperk\"ahler quotient construction, due to Hitchin, Karlhede, Lindstrom and Rocek \cite{HKLR} carries an anti-self dual conformal structure, allowing Penrose's Twistor theory \cite{twistor} techniques to be applied in this case.

Recently works in emergent gravity \cite{Lee:2012px} aim at constructing a Riemannian geometry from $U(1)$ gauge fields on a noncommutative spacetime. This construction is invertible to find corresponding $U(1)$ gauge fields on a (generalized) Poisson manifold given a metric $(M, g)$. There are already detailed tests \cite{Lee:2012rb} of the emergent gravity picture with explicit solutions in both gravity and gauge theory sides. Symplectic U(1) gauge fields have been derived starting from the Eguchi-Hanson metric in four-dimensional Euclidean gravity. The result precisely reproduces the U(1) gauge fields of the Nekrasov-Schwarz instanton previously derived from the top-down approach. To clarify the role of noncommutative spacetime, the prescription has been inverted and Braden-Nekrasov $U(1)$ instanton defined in ordinary commutative spacetime was used to derive a corresponding gravitational metric just to show that the K\"ahler manifold determined by the Braden-Nekrasov instanton exhibits a spacetime singularity while the Nekrasov-Schwarz instanton gives rise to a regular geometry-in the form of Eguchi-Hanson space. This result implies the important role noncommutativity of spacetime plays in resolving spacetime singularities \cite{Lee:2012ju} in general relativity. Some relevant studies related to emergent nature of Schwarzschild spacetime was also performed in \cite{CGR}.

One may wonder if we can get $U(1)$ gauge fields in the same way from the Taub-NUT metric.
A critical difference from the Eguchi-Hanson metric \cite{egh-report}  is that the Taub-NUT metric (\ref{tn})  is
locally asymptotic at infinity to $\mathbb{R}^3 \times \mathbb{S}^1$, and so it belongs to the class of asymptotically locally flat (ALF) spaces. Thus, the Hopf coordinates cannot represent the Taub-NUT metric, and it is difficult to naively generalize the same construction to ALF spaces. From gauge theory perpective, it may be related to the fact that ALF spaces arise from NC monopoles \cite{lee-yi} whose underlying equation is defined by an $\mathbb{S}^1$-compactification of  the self(anti)-dual-instanton equation, the so-called Nahm equation. We will discuss in \cite{future-paper} a possible generalization to include the Taub-NUT in the bottom-up approach of emergent gravity.

It is known that only for special choice of the NUT parameter we get a regular metric, but generally, one encounters singularities at either end point of the 4-dim radial coordinate. In the most generic case, although with a particular choice of the period for the azimuthal angle, one can get away with the bolt-singularity, the NUT singularity (co-dimension 4 orbifold singularity) stays, possibly admitting an M theory interpretation associated with the corresponding non-abelian gauge symmetries \cite{KB}.

Recently, Ricci flat metrics of ultrahyperbolic signature were constructed \cite{tomsk} with l-conformal Galilei symmetry, involving an $AdS_2$ part reminiscent of the near horizon geoemtry of extremal black holes. Similarly, it should be interesting to see if Taub-NUT spaces are associable with geodesics that can describe second order dynamical systems. Perhaps the most interesting issue will be to explore whether something like ``Taub-NUT/CFT'' correspondence can be conjectured.

\vspace{0.2cm}

\noindent

\newpage
{\bf Acknowledgements} The research of RR was supported by FAPESP through Instituto de 
Fisica, Universidade de Sao Paulo with grant number 2013/17765-0. This work was performed
during RR's visit to S.N. Bose National Centre for Basic Sciences in Kolkata. He thanks SNBNCBS 
for the hospitality and support during that period.

\section{APPENDIX}

Important computations and derivations of this article are provided in this section.

\numberwithin{equation}{subsection}

\subsection{Basic Killing tensors from Holten's Algorithm}

\subsubsection*{Angular Momentum}

If we choose to set $C^{(n)}_{\{ i \}} = 0, \hspace{0.25cm} \forall \hspace{0.25cm} n \geq 2$, we get the Killing equations:

\begin{equation}
\label{fcon1} \nabla_{(i} C^{(1)}_{j)} = 0
\end{equation}

There are two parts of this solution we shall study in detail. We can write (\ref{fcon1}) as:

\begin{equation}
\nabla_i C^{(1)}_j + \nabla_j C^{(1)}_i = 0 \hspace{0.75cm} \Rightarrow \hspace{0.75cm} \nabla_i C^{(1)}_j = - \nabla_j C^{(1)}_i
\end{equation}

This is an anti-symmetric matrix, written as $\theta_{ij} = - \theta_{ji}$. Further elaboration gives:

\vspace{-0.5cm}
\[ \begin{split}
\theta_{ij} (\vec{x}) = \varepsilon_{ijk} (\vec{x}) \theta^k &= g_{im} (\vec{x}) {\varepsilon^m}_{jk} \theta^k \\
\therefore \hspace{0.5cm} - \nabla_j C^{(1)}_i = g_{im} (\vec{x}) {\varepsilon^m}_{jk} \theta^k \hspace{0.5cm} &\Rightarrow \hspace{0.5cm} C^{(1)}_i = - g_{im} (\vec{x}) {\varepsilon^m}_{jk} \theta^k x^j
\end{split} \]

Thus, we have the rotation operator as the 1st order co-efficient:

\begin{equation}
\label{con1} C^{(1)}_i = - g_{im} (\vec{x}) {\varepsilon^m}_{jk} \theta^k x^j
\end{equation}

Applying this co-efficient into the 1st term of the power series, we get:

\vspace{-0.5cm}
\[ \begin{split}
Q^{(1)} &= C_i^{(1)} \Pi^i = - g_{im} (\vec{x}) {\varepsilon^m}_{jk} \theta^k x^j \Pi^i \\ 
\Rightarrow \hspace{1cm} & \bm{L} . \bm{\theta} = - \big( \varepsilon_{ijk} \Pi^i x^j \big) \theta^k = \big( \bm{x} \times \bm{\Pi} \big) . \bm{\theta}
\end{split} \]

\begin{equation}
\therefore \hspace{1cm} \bm{L} = \bm{x} \times \bm{\Pi}
\end{equation}

This eventually becomes the conserved quantity known as the angular momentum.

\subsubsection*{Laplace-Runge-Lenz vector}

Now when we choose to set $C^{(n)}_{\{ i \}} = 0, \hspace{0.25cm} \forall \hspace{0.25cm} n \geq 3$, we get the Killing equations:

\begin{equation}
\label{fcon2} \nabla_i C^{(2)}_{jk} + \nabla_j C^{(2)}_{ki} + \nabla_k C^{(2)}_{ij} = 0
\end{equation}

as we can see, (\ref{fcon2}) perfectly matches the property of the Killing Yano and Killing St$\ddot{\text{a}}$ckel tensors. The Runge-Lenz like quantity is given by a symmetric sum as shown below:

\vspace{-0.5cm}
\begin{align} 
\Big[ \vec{A} \times \big( \vec{B} \times \vec{C} \big) \Big]_i &= \varepsilon_{ilm} {\varepsilon^m}_{jk} A^l B^j C^k \hspace{1cm} \varepsilon_{ilm} {\varepsilon^m}_{jk} = \delta_{ij} \delta_{lk} - \delta_{ik} \delta_{lj} \\ \nonumber\\
\nabla_k C^{(2)}_{ij} &= \varepsilon_{i l m} (\vec{x}) {\varepsilon^m}_{jk} (\vec{x}) n^l + (i \leftrightarrow j) \nonumber \\  \displaybreak[0]
&= \big( 2 g_{ij} (\vec{x}) g_{kl} (\vec{x}) - g_{ik} (\vec{x}) g_{jl} (\vec{x}) - g_{il} (\vec{x}) g_{kj} (\vec{x}) \big) n^l x^k \nonumber \\ \nonumber \\
\label{con2} \therefore \hspace{0.5cm} C^{(2)}_{ij} &= \big( 2 g_{ij} (\vec{x}) n_k - g_{ik} (\vec{x}) n_j - g_{kj} (\vec{x}) n_i \big) x^k
\end{align}

As before, applying this co-efficient to the 2nd  order term in the power series gives

\vspace{-0.5cm}
\[ \begin{split}
Q^{(2)} &= \frac12 C^{(2)}_{ij} \Pi^i \Pi^j = \big\{ \big| \bm{\Pi} \big|^2 \big( \bm{n} . \bm{x} \big) - \big( \bm{\Pi} . \bm{x} \big) \big( \bm{\Pi} . \bm{n} \big) \big\} \\ 
&= \bm{N} . \bm{n} = \big\{ \big| \bm{\Pi} \big|^2 \bm{x} - \big( \bm{\Pi} . \bm{x} \big) \bm{\Pi} \big\} . \bm{n} = \big\{ \bm{\Pi} \times \big( \bm{x} \times \bm{\Pi} \big) \big\} . \bm{n}
\end{split} \]

\begin{equation}
\therefore \hspace{1cm} \bm{N} = \bm{\Pi} \times \big( \bm{x} \times \bm{\Pi} \big)
\end{equation}

This quantity is a term that is present in another conserved quantity known as the Laplace-Runge-Lenz vector. Having found the two familiar types of conserved quantities, we can now proceed to see what it looks like for the Taub-NUT metric.

\subsection{The Bohlin transformation}

The Bohlin transformation that maps the co-ordinate system on a plane, is given by:

\vspace{-0.25cm}
\begin{equation}
f: z \longrightarrow \xi^\alpha = \big( z^\alpha \big)^2 = R e^{i \phi} \hspace{0.5cm} \Rightarrow \hspace{0.5cm} z = \xi^{\frac 1 2}
\end{equation}

Now we must note that another Noether invariant, the angular momentum will change form under this transformation. We re-parametrize to preserve the form of angular momentum.

\vspace{-0.5cm}
\begin{align}
l = r^2 \dot{\theta} = \vert z \vert ^2 \dot{\theta} = \vert \xi \vert^2 \phi' \hspace{0.5cm} &\Rightarrow \hspace{0.5cm} \vert \xi \vert \frac{d \tilde{\tau}}{d \tau} \theta' = \vert \xi \vert^2 \theta'\\
 \therefore \hspace{0.5cm} \tau \longrightarrow \tilde{\tau} : \frac{d \tilde{\tau}}{d \tau} &= \vert \xi \vert
\end{align}

The velocity and acceleration can be given as:

\vspace{-0.5cm}
\begin{align}
\dot{z}^\alpha &= \frac 1 2 \frac{\vert \xi \vert}{\big( \xi^\alpha \big)^{\frac 1 2}} {\xi^\alpha}' = \frac 1 2 \big( \bar{\xi}^\alpha \big)^{\frac 1 2} {\xi^\alpha}'\\
\ddot{z}^\alpha &= \frac 1 2 \vert \xi \vert \frac d {d \tilde{\tau}} \Big\{ \big( \bar{\xi}^\alpha \big)^{\frac 1 2} {\xi^\alpha}' \Big\} = \frac 1 2 \frac{\vert \xi \vert^2}{\big( \xi^\alpha \big)^{\frac 1 2}} {\xi^\alpha}'' + \frac 1 4 \big( \xi^\alpha \big)^{\frac 1 2} \vert {\xi}' \vert^2
\end{align}

The equation of motion for a Harmonic Oscillator eventually becomes:

\vspace{-0.5cm}
\begin{align} \displaybreak[0]
m \bigg\{ \frac 1 2 \frac{\vert \xi \vert^2}{\big( \xi^\alpha \big)^{\frac 1 2}} {\xi^\alpha}'' + \frac 1 4 \big( \xi^\alpha \big)^{\frac 1 2} \vert {\xi}' \vert^2 \bigg\} &= - k \big(\xi^\alpha\big)^{\frac 1 2} \nonumber \\
\Rightarrow \hspace{0.5cm} \vert \xi \vert^2 {\xi^\alpha}'' + \frac 1 2 \xi^\alpha \vert {\xi}' \vert^2 = - \frac {2k} m \xi^\alpha \hspace{0.5cm} &\Rightarrow \hspace{0.5cm} {\xi^\alpha}'' = - \bigg( \frac 1 2 \vert {\xi}' \vert^2 + \ \frac {2k} m \bigg) \frac{\xi^\alpha}{\vert \xi \vert^2}
\end{align}

The Hamiltonian $\mathcal{H}$ of the oscillator can be re-written to complete the transformation:

\vspace{-0.5cm}
\begin{align}
\mathcal{H} &= \frac m 2 \vert \dot{z} \vert^2 + \frac k 2 \vert z \vert^2 = \frac m 4 \bigg( \frac 1 2 \vert {\xi'} \vert^2 + \frac {2k} m \bigg) \vert \xi \vert \hspace{1cm} \Rightarrow \hspace{1cm} \bigg( &\frac {\vert \xi' \vert^2 } 2 + \frac {2k} m \bigg) = \frac{4 \mathcal{H}} m \frac 1 {\vert \xi \vert} = \kappa \frac 1 {\vert \xi \vert} \nonumber \\ 
& \hspace{2cm} \therefore \hspace{1cm} {\xi^\alpha}'' = - \bigg( \frac {\vert {\xi^\alpha}' \vert^2 } 2 + \frac {2k} m \bigg) \frac {\xi^\alpha}{\vert \xi \vert^2} \equiv - \kappa \frac {\xi^\alpha}{\vert \xi \vert^3}
\end{align}

showing that it  restores the central force nature of the system, giving us the equation of motion for inverse square law forces.

\subsection{Double derivative of Killing-Yano tensors}

Similar to Killing vectors, rank n Killing-Yano tensors exhibit a curvature equation

$$\big( \nabla_a \nabla_b - \nabla_b \nabla_a \big) K_{c_1 . . . c_n} = \sum_{i = 1}^n {R_{a b c_i}}^d K_{c_1 . . . d . . . c_n}$$
For the LHS, by permuting the indices according to the rules, we will get

\vspace{-0.5cm}
\[ \begin{split}
\text{LHS } =  \big( \nabla_a \nabla_b - \nabla_b \nabla_a \big) K_{c_1 . . . c_n} &= - \nabla_a \nabla_{c_1} K_{b c_2 . . . c_n} + \nabla_b \nabla_{c_1} K_{a c_2 . . . c_n} \\
= 2 \nabla_{c_1} \nabla_b K_{a c_2 . . . c_n} &- {R_{a b c_1}}^d K_{d c_2 . . . c_n} + \sum_{i = 2}^n \big( {R_{b c_1 c_i}}^d K_{a c_2 . . . d . . . c_n} - {R_{a c_1 c_i}}^d K_{b c_2 . . . d . . . c_n} \big) \\
& = {R_{a b c_1}}^d K_{d c_2 . . . c_n} + \sum_{i = 2}^n {R_{a b c_i}}^d K_{c_1 . . . d . . . c_n}
\end{split} \]

\vspace{-0.5cm}
\[ \begin{split}
2 \nabla_{c_1} \nabla_b K_{a c_2 . . . c_n} = 2 {R_{a b c_1}}^d K_{d c_2 . . . c_n} + &\sum_{i = 2}^n \big( \underbrace{{R_{a c_1 c_i}}^d K_{b c_2 . . . d . . . c_n} - {R_{b c_1 c_i}}^d K_{a c_2 . . . d . . . c_n}}_\text{II} + \underbrace{{R_{a b c_i}}^d K_{c_1 . . . d . . . c_n}}_\text{I} \big) \\
\because \hspace{1cm} \nabla_{c_1} \nabla_b K_{a c_2 . . . c_n} = \nabla_{c_1} \nabla_{[b} &K_{a c_2 . . . c_n]}, \hspace{1cm} W_{c_1} := \nabla_{c_1} \nabla_{[b} K_{a c_2 . . . c_n]} e^a \wedge e^b \wedge e^{c_2} . . . \wedge e^{c_n}
\end{split} \]
With I and II above, we can create the 3-forms

\vspace{-0.5cm}
\[ \begin{split}
\text{I} \hspace{0.5cm} \rightarrow \hspace{0.5cm} {R_{a b c_i}}^d e^a \wedge e^b \wedge e^{c_i} = \frac13 \big( {R_{a b c_i}}^d &+ {R_{b c_i a}}^d + {R_{c_i a b}}^d \big) e^a \wedge e^b \wedge e^{c_i} = 0 \\ 
\text{II} \hspace{0.5cm} \rightarrow \hspace{0.5cm} {R_{a c_1 c_i}}^d K_{b c_2 . . . d . . . c_n} e^a \wedge e^b \wedge e^{c_i} &= - {R_{c_1 a c_i}}^d K_{b c_2 . . . d . . . c_n} e^a \wedge e^b \wedge e^{c_i} \\
&= - {R_{c_1 c_i b}}^d K_{a c_2 . . . d . . . c_n} e^a \wedge e^b \wedge e^{c_i}
\end{split} \]
Thus, II will become:

\vspace{-0.5cm}
\[ \begin{split}
- \sum_{i = 2}^n \big( {R_{c_1 c_i b}}^d K_{a c_2 . . . d . . . c_n} + {R_{b c_1 c_i}}^d K_{a c_2 . . . d . . . c_n} \big) &e^a \wedge e^b \wedge e^{c_i} = \sum_{i = 2}^n {R_{c_i b c_1}}^d K_{a c_2 . . . d . . . c_n} e^a \wedge e^b \wedge e^{c_i} \\
= \sum_{i = 2}^n {R_{a b c_1}}^d K_{d c_2 . . . c_i . . . c_n} &e^a \wedge e^b \wedge e^{c_i} = ( n - 1 ) {R_{a b c_1}}^d K_{d c_2 . . . c_i . . . c_n} e^a \wedge e^b \wedge e^{c_i}
\end{split} \]
Applying this result back in the main equation, we get:

\vspace{-0.5cm}
\[ \begin{split}
\therefore \hspace{1cm} 2 \nabla_{c_1} \nabla_b K_{a c_2 . . . c_n}  e^a \wedge e^b \wedge e^{c_i} &= \big( 2 {R_{a b c_1}}^d K_{d c_2 . . . c_n} + ( n - 1 ) {R_{a b c_1}}^d K_{d c_2 . . . c_n} \big) e^a \wedge e^b \wedge e^{c_i} \\
\Rightarrow \hspace{1cm} 2 \nabla_{c_1} \nabla_b K_{a c_2 . . . c_n}  e^a \wedge e^b \wedge e^{c_i} &= ( n + 1 ) {R_{a b c_1}}^d K_{d c_2 . . . c_n} e^a \wedge e^b \wedge e^{c_i}
\end{split} \]
Finally, we get the double-derivative of KY tensors as:

\begin{equation}
\label{dderiv} \therefore \hspace{1cm} \nabla_a \nabla_b K_{c_1 c_2 . . . c_n} = ( - 1 )^{n + 1} \frac{n+1}2 {R_{[bc_1|a|}}^d K_{c_2 c_3 . . . c_n ] d}
\end{equation}

\end{document}